\documentclass[12pt, draftclsnofoot, onecolumn]{IEEEtran}

\usepackage{graphicx}
\usepackage{amsmath,bbm,epsfig,amssymb,amsfonts,amstext,verbatim,amsopn,cite,subfigure,multirow,multicol,lipsum}
\usepackage{balance}
\usepackage{url}
\usepackage{amsfonts}
\usepackage{epsfig}
\usepackage{setspace}
\usepackage{stmaryrd}
\usepackage{psfrag}	
\usepackage{multirow}
\usepackage{float}
\usepackage[process=auto]{pstool}
\usepackage{etoolbox}
\usepackage{algorithm}
\usepackage{algorithmic}
\usepackage{hyperref}

\usepackage{pifont}
\allowdisplaybreaks

\newcommand*{\Scale}[2][4]{\scalebox{#1}{$#2$}}%
\newcommand*{\QEDwhite}{\hfill\ensuremath{\square}}%

\newtheorem{theo}{Theorem}
\newtheorem{lem}{Lemma}
\newtheorem{remk}{Remark}
\newtheorem{defin}{Definition}
\newtheorem{prop}{Proposition}
\newtheorem{corol}{Corollary}
\newtheorem{examp}{Example}

\newtoggle{ConfVersion}

\toggletrue{ConfVersion}

\iftoggle{ConfVersion}{%
  
}{%
  
}

\EndPreamble
\begin{document}

\title{\hspace{-0.5cm} SCW Codes for Maximum Likelihood Detection \hspace{-0.5cm}\\
in Diffusive Molecular Communications \\
without Channel State Information \vspace{-0.3cm}}

\author{Vahid Jamali\dag, Arman Ahmadzadeh\dag, Nariman Farsad\ddag, and Robert Schober\dag\\
\IEEEauthorblockA{\dag Friedrich-Alexander University (FAU), Erlangen, Germany  \\
\ddag Stanford University, Stanford, California, USA \vspace{-0.3cm}
\thanks{This paper has been accepted in part for presentation at IEEE ISIT 2017 \cite{ISIT17_IEEE}.}  
\thanks{This work was supported in part by the German Science Foundations (Project SCHO 831/7-1) and the Friedrich-Alexander-University Erlangen-N\"urnberg under the Emerging Fields Initiative (EFI).}
}
}

\maketitle

\begin{abstract}
Instantaneous or statistical channel state information (CSI) is needed for most detection schemes developed  for molecular communication (MC) systems. Since the MC channel changes over time, e.g., due to variations in the velocity of  flow, the temperature, or the distance between transmitter and receiver,  CSI acquisition has to be conducted repeatedly to keep track of CSI variations. Frequent CSI acquisition  may entail a large overhead whereas infrequent CSI acquisition may result in a low CSI estimation accuracy. To overcome these challenges,  we design codes  which enable maximum likelihood sequence detection at the receiver without \textit{instantaneous} or \textit{statistical} CSI. In particular, assuming concentration shift keying modulation, we show that a class of codes, referred to as \emph{strongly constant-weight (SCW) codes}, enables \textit{optimal CSI-free} sequence detection at the expense of a decrease in data rate. For the proposed SCW codes, we analyze the code rate, the error rate, and the average number of released molecules. In addition, we study the properties of  binary SCW codes and balanced~SCW~codes in further detail. Simulation results verify our analytical derivations and reveal that SCW codes with CSI-free detection outperform uncoded transmission with optimal coherent and non-coherent detection.
\end{abstract}


\section{Introduction}

In contrast to conventional wireless communication systems that encode data into electromagnetic waves, synthetic molecular communication (MC) systems are envisioned to embed data into the characteristics of  signaling molecules such as their concentration, type, and time of release  \cite{Nariman_Survey,Survey_Mol_Net}. Diffusive MC  is a common  strategy for communication between nano-/microscale entities in nature such as bacteria, cells, and organelles (i.e., components of cells) \cite{CellBio,BioPhysic}. Therefore, diffusive MC has been considered as a bio-inspired approach for communication between small-scale nodes for applications where conventional wireless communication may be inefficient or even infeasible \cite{Nariman_Survey,Survey_Mol_Nono}.

\subsection{Motivation}

In diffusive MC, the expected number of signalling molecules observed at the receiver at a given time after the emission of a known number of molecules by the transmitter and the expected number of interfering molecules observed at the receiver constitute the channel state information (CSI) \cite{HamidJSAC,TCOM_MC_CSI,NanoCOM16}. Knowledge of the \textit{instantaneous} CSI is needed in general for optimal coherent detection \cite{HamidJSAC} and can be obtained using  training sequence-based channel estimators \cite{TCOM_MC_CSI}. The CSI of an MC channel depends on  various parameters such as the diffusion coefficient of the signaling molecules, the velocity of the flow in the channel, the concentration of enzyme degrading the signaling molecules, the distance between the transmitter and the receiver, etc., see \cite[Chapter~4]{BioPhysic}, \cite[Chapters~3 and 4]{Berg}, \cite{ArmanMobileMC,ArmanMCStat}. A change in any of these parameters affects the CSI of the considered MC channel. Therefore, CSI acquisition has to be  conducted repeatedly to keep track of CSI variations.  To reduce the CSI acquisition overhead, the authors in \cite{NanoCOM16} derived the optimal non-coherent detector which requires only \textit{statistical} CSI instead of instantaneous CSI. The statistical CSI of a particular MC channel can be estimated using empirical measurements. However, this may not always be  possible, especially not for practical MC systems with limited processing capabilities. In fact,  an experimentally verified statistical channel model for MC systems has not been reported yet. Motivated by the aforementioned challenges in CSI acquisition, the goal of this paper is to design codes which enable optimal detection without CSI at the receiver.

\subsection{Contributions}

In this paper, we consider concentration shift keying (CSK) modulation, where information is encoded in the number of molecules released by the transmitter, and formulate the maximum likelihood (ML) problem  for both coherent and non-coherent sequence detection. The coherent and non-coherent ML sequence detectors require in general instantaneous and statistical CSI, respectively.  However, based on the intuition obtained from the structure of the optimal detectors,  we propose a class of codes, referred to as \textit{strongly constant-weight (SCW) codes}, for which ML detection is possible without instantaneous or statistical CSI knowledge. In other words, SCW codes enable optimal CSI-free detection at the expense of a decrease in data rate. For the  proposed SCW codes, we analyze the code rate, the error rate, and the average number of released molecules. In addition, we study the properties of  binary SCW codes and balanced~SCW~codes in further detail. Simulation results verify our analytical derivations and reveal that SCW codes with CSI-free detection outperform uncoded transmission with optimal coherent and non-coherent detection.

\subsection{Related Work}

We note that the problem considered in this paper, i.e., the design of SCW codes, can be seen as a modulation design or coded modulation design problem \cite{Nariman_Survey,Huber_Polar,Ungerboeck_CodedMod}. In fact, the SCW codewords in the codebook can be seen as symbols (hyper-symbols) in a corresponding multi-dimensional symbol consellation. Various modulation techniques have been proposed so far for MC systems, see \cite{Nariman_Survey} for a comprehesive overview. For instance, the widely-adopted on-off keying (OOK) modulation is a special case of CSK modulation where for binary one and zero, $N^{\mathrm{tx}}$ and zero molecules are released by the transmitter, respectively \cite{ConsCIR,OOK_MC}.  Information can be also encoded in the time of release of molecules \cite{Nariman_Timing}. A special case is pulse position modulation (PPM) where data is encoded in the time at which molecules are released by the transmitter to form a pulse \cite{PPM_Pieroborn}. We note that optimal detection for the modulation techniques proposed in \cite{ConsCIR,OOK_MC,Nariman_Timing,PPM_Pieroborn} generally requires instantaneous CSI of the MC channel. In fact, only for the special case of binary PPM, it has been shown that knowledge of CSI is not needed for optimal detection in an inter-symbol interference (ISI)-free MC channel \cite{PPM_Pieroborn}. As we show in this paper, the proposed SCW codes include PPM as a special case when interpreting codewords as hyper-symbols.

Coded modulation has been extensively studied for conventional wireless communications \cite{Huber_Polar,Ungerboeck_CodedMod,Huber_MultiLevCode}. Thereby,  coded modulation is typically adopted to enhance reliability especially for large symbol constellations.  However, in this paper, our main motivation for employing SCW codes is to devise an optimal ML detection algorithm that does not require CSI. We note that SCW codes  are a special case of the widely-known constant-weight (CW) codes \cite{WeightCodeClass,CW_qray}. In fact, CW codes have been extensively investigated in the literature, see e.g. \cite{WeightCodeClass} for binary CW codes, \cite{CW_qray} for $q$-ary CW codes, \cite{BalancedCode_Knuth} for balanced codes, \cite{MutiplyCW} for multiply CW codes, etc. Moreover, multiple pulse position modulation (MPPM) was developed for optical communications and constitutes a special case of the proposed SCW codes \cite{MPPM_Optic,MPPM_Rate}.   However, to the best of the authors' knowledge, SCW codes and the ensuing CSI-free detection have not been considered in the literature,~yet.

\subsection{Organization and Notation}

The remainder of this paper is organized as follows. In Section II, the system model adopted in this paper is presented. In Section~III, we first provide the optimal coherent and non-coherent detectors for general transmit sequences. Subsequently, we introduce the SCW codes and derive the corresponding optimal CSI-free detector. In Section~IV, the code rate, error rate, and average number of released molecules of the proposed SCW codes are analyzed. Numerical results are presented in Section~V, and conclusions are drawn in Section~VI.

\textit{Notations:} We use the following notations throughout this paper: $\mathsf{E}\{x\}$ and $\mathsf{Var}\{x\}$ denote the expectation and the variance of random variable (RV) $x$. Bold lower case letters  denote vectors and $\mathbf{a}^{\mathsf{T}}$ represents the transpose of vector $\mathbf{a}$. $H_n(\cdot)$ represents the entropy function for the logarithm to base $n$, $n!$ is the factorial of $n$, and $O(n)$ denotes the complexity order of $n$.  Moreover,  $\mathcal{P}(\lambda)$  denotes a Poisson RV with mean  $\lambda$, $\lfloor\cdot\rfloor$ denotes the floor function which maps a real number to the largest integer number that is smaller or equal to the real number, and $\mathbf{1}\{\cdot\}$ is an indicator function that is equal to one if the argument is true and equal to zero otherwise.

\section{System Model}

\begin{figure}
  \centering
 \scalebox{0.6}{
\pstool[width=1\linewidth]{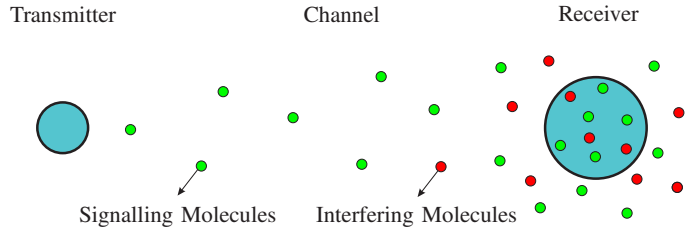}{
\psfrag{T}[c][c][1.2]{Transmitter}
\psfrag{R}[c][c][1.2]{Receiver}
\psfrag{C}[c][c][1.2]{Channel}
\psfrag{S}[c][c][1.2]{Signalling Molecules}
\psfrag{N}[c][c][1.2]{Interfering Molecules}
} } \vspace{-0.3cm}
\caption{Schematic illustration of the considered MC system.\vspace{-0.3cm}}
\label{Fig:TxRx}
\end{figure}

We consider an MC system  consisting of a transmitter, a channel, and a receiver, see Fig.~\ref{Fig:TxRx}. We employ CSK modulation  where the transmitter releases  $s[k]N^{\mathrm{tx}}$ molecules at the beginning of the $k$-th symbol interval to convey symbol $s[k]\in\mathcal{S}$ \cite{Nariman_Survey}. Here, $N^{\mathrm{tx}}$ is the maximum number of molecules that the transmitter can release in one symbol interval, i.e., a peak per-symbol ``power" constraint is employed, and $\mathcal{S}=\{\eta_0,\eta_1,\dots,\eta_{L-1}\}$ denotes the symbol set  where $L$ is the number of available symbols. Without loss of generality, we assume $\eta_0<\eta_1<\cdots<\eta_{L-1}$, $\eta_0=0$, and $\eta_{L-1}=1$. Moreover, let $\mathbf{s}=[s[1],s[2],\dots,s[K]]^{\mathsf{T}}$  denote a codeword  comprising $K$ symbols. 

The released molecules diffuse through the fluid medium between the transmitter and the receiver. We assume that the movements of individual molecules are  independent from each other. The number of observed (counted) molecules at the receiver in each symbol interval constitutes the received signal. Let $\mathbf{r}=[r[1],r[2],\dots,r[K]]^{\mathsf{T}}$ denote the vector of observations corresponding to sequence $\mathbf{s}$ where  $r[k]$ denotes the number of  molecules observed at the receiver  in symbol interval $k$. Due to the counting process at the receiver,  $r[k]$ can be accurately modelled as a Poisson RV\footnote{We note that $r[k]$ is exactly modelled by as a bionomial RV \cite{NanoCOM16,Yilmaz_Poiss,HamidJSAC}. However, since the binomial distribution makes analysis difficult, $r[k]$ is often approximated by  Poisson or Gaussian models. For instance,  using the analytical framework developed in \cite{Yilmaz_Poiss}, it can be shown that for $N^{\mathrm{tx}}=1000$, if $p_{\mathrm{s}}\leq 0.115$ holds, the Poisson distribution more accurately approximates the binomial distribution  in terms of the root mean squared error (RMSE) of the cumulative distribution function (CDF), whereas, if $p_{\mathrm{s}}>0.115$ holds, the Gaussian approximation is a better fit. For typical MC systems, if  $N^{\mathrm{tx}}=1000$ molecules are released by the transmitter, we expect to observe much fewer than $N^{\mathrm{tx}}p_{\mathrm{s}}=113$ molecules at the receiver. Hence, we adopt the Poisson approximation in this paper as it is more accurate compared to the Gaussian approximation~for~typical MC~applications.}, see \cite{NanoCOM16,Yilmaz_Poiss,HamidJSAC}, i.e.,
\begin{align} \label{Eq:ChannelInOut}
  r[k]  \sim \mathcal{P}(s[k]\bar{c}_{\mathrm{s}}+\bar{c}_{\mathrm{n}}), 
\end{align}
where   $\bar{c}_{\mathrm{s}}$ is the  number of  molecules \textit{expected} to be observed at the receiver  in symbol interval $k$ due to the release of $N^{\mathrm{tx}}$ molecules by the transmitter at the beginning of symbol interval~$k$ and  $\bar{c}_{\mathrm{n}}$ is the \textit{expected} number of interfering noise molecules comprising  multiuser interference (caused by other MC links) and  external noise (originating from natural sources) observed by the receiver \cite{NanoCOM16}. The inter-symbol interference (ISI) free communication model in (\ref{Eq:ChannelInOut}) implies that the symbol duration is chosen large enough such that the channel impulse response (CIR) approaches zero at the end of a symbol interval.
We note that enzymes \cite{Adam_Enzyme} and reactive information molecules, such as acid/base molecules \cite{Acid_Base}, may be used to shorten the CIR.

The channel model in (\ref{Eq:ChannelInOut}) implicitly includes both diffusion noise and interference. To explicitly distinguish the signal, the noise, and the interference terms, we rewrite (\ref{Eq:ChannelInOut}) as \cite{CL_MF_IEEE}
\begin{align} \label{Eq:SigNoiseInt}
  r[k]  =  \underset{\text{deterministic}}{\underbrace{\overset{\text{signal}}{\overbrace{s[k]\bar{c}_{\mathrm{s}}}} + \overset{\text{constant}}{\overbrace{\bar{c}_{\mathrm{n}}}}}} 
  + \underset{\text{random}}{\underbrace{\overset{\text{noise}}{\overbrace{n[k]}} + \overset{\text{interference}}{\overbrace{I[k]}}}}.
\end{align}
Here, $n[k]$ is a noise term with  mean zero and variance $s[k]\bar{c}_{\mathrm{s}}$, i.e., the signal-dependent diffusion noise, and $I[k]$ is the   interference term with  mean zero and variance  $\bar{c}_{\mathrm{n}}$. In fact, $n[k]$ and $I[k]$ are RVs equivalent to Poisson RVs whose means are subtracted, i.e., $n[k]\sim\big(\mathcal{P}(s[k]\bar{c}_{\mathrm{s}})-s[k]\bar{c}_{\mathrm{s}}\big)$ and $I[k]\sim\big(\mathcal{P}(\bar{c}_{\mathrm{n}})-\bar{c}_{\mathrm{n}}\big)$. For future reference, we define  $\mathsf{SIR}=\frac{\bar{c}_{\mathrm{s}}}{\bar{c}_{\mathrm{n}}}$ as the signal-to-interference ratio (SIR) and $\mathsf{SINR}=\frac{\left(\mathsf{E}\{s[k]\bar{c}_{\mathrm{s}}\}\right)^2}{\mathsf{Var}\{r[k]\}}\big |_{s[k]=1} =\frac{\bar{c}_{\mathrm{s}}^2}{\bar{c}_{\mathrm{s}}+\bar{c}_{\mathrm{n}}}$ as the signal-to-interference-plus-noise ratio (SINR).

Note that the MC channel in (\ref{Eq:ChannelInOut}) is characterized by $\bar{c}_{\mathrm{s}}$ and $\bar{c}_{\mathrm{n}}$. Hence, we refer to  vector $\bar{\mathbf{c}}=[\bar{c}_{\mathrm{s}},\bar{c}_{\mathrm{n}}]^{\mathsf{T}}$ as the CSI of the considered MC system in the remainder of this paper. Moreover, we assume that the CSI remains unchanged over one block of transmitted symbols, i.e., one codeword, but may change from one block to the next (e.g., due to a change in the flow velocity or the distance between transmitter and receiver). To model this, we assume that the CSI, $\bar{\mathbf{c}}$, is an RV that takes its values in each block according to probability density function (PDF)~$f_{\bar{\mathbf{c}}}(\bar{c}_{\mathrm{s}},\bar{c}_{\mathrm{n}})$.

\section{Optimal CSI-Free Detection Using SCW Codes}

In this section, we first formulate the ML problems for coherent and non-coherent sequence detection which in general require instantaneous and statistical CSI, respectively. Subsequently, we introduce the SCW codes for which we derive a CSI-free ML sequence detector. 

\subsection{Coherent and Non-Coherent ML Sequence Detection}

The  ML problems for coherent and non-coherent sequence detection can be mathematically formulated as
\begin{align} 
  \hat{\mathbf{s}}^{\mathrm{c}}  &= 
  \underset{\mathbf{s}\in\boldsymbol{\mathcal{S}}}{\mathrm{argmax}} \,\, f_{\mathbf{r}}(\mathbf{r}|\bar{\mathbf{c}},\mathbf{s}) \quad \text{and}
   \label{Eq:ML}\\
   \hat{\mathbf{s}}^{\mathrm{nc}}  &= \underset{\mathbf{s}\in\boldsymbol{\mathcal{S}}}{\mathrm{argmax}} \,\, 
     \int_{\bar{c}_{\mathrm{s}}} \int_{\bar{c}_{\mathrm{n}}} f_{\mathbf{r}}(\mathbf{r}|\bar{\mathbf{c}},\mathbf{s}) f_{\bar{\mathbf{c}}}(\bar{c}_{\mathrm{s}},\bar{c}_{\mathrm{n}}) \mathrm{d}\bar{c}_{\mathrm{s}} \mathrm{d}\bar{c}_{\mathrm{n}}, \label{Eq:ML_NonCoherent}
\end{align}
respectively, where $\boldsymbol{\mathcal{S}}$ is the set of available sequences $\mathbf{s}$ and $f_{\mathbf{r}}(\mathbf{r}|\bar{\mathbf{c}},\mathbf{s})$ is the PDF of received vector $\mathbf{r}$ conditioned on a given CSI vector, $\bar{\mathbf{c}}$, and a given hypothesis sequence $\mathbf{s}$. Exploiting the fact that the observations in different symbol intervals are independent, we obtain $f_{\mathbf{r}}(\mathbf{r}|\bar{\mathbf{c}},\mathbf{s})$ as 
\begin{align} 
f_{\mathbf{r}}(\mathbf{r}|\bar{\mathbf{c}},\mathbf{s}) = \prod_{k=1}^{K} \frac{\left( \bar{c}_{\mathrm{s}} s[k] + \bar{c}_{\mathrm{n}} \right)^{r[k]} \mathsf{exp}\left(- \bar{c}_{\mathrm{s}} s[k] - \bar{c}_{\mathrm{n}} \right)}{r[k]!}. \label{Eq:Likelihood}
\end{align}
For general sets $\boldsymbol{\mathcal{S}}$, for coherent ML sequence detection, instantaneous CSI, i.e., $(\bar{c}_{\mathrm{s}},\bar{c}_{\mathrm{n}})$, is required, cf. (\ref{Eq:ML}) and (\ref{Eq:Likelihood}), whereas for non-coherent ML sequence detection, statistical CSI, i.e., $f_{\bar{\mathbf{c}}}(\bar{c}_{\mathrm{s}},\bar{c}_{\mathrm{n}})$, is required, cf. (\ref{Eq:ML_NonCoherent}) and (\ref{Eq:Likelihood}).

In the following, we simplify (\ref{Eq:ML}) to facilitate the development of the proposed CSI-free detector in the next subsection\footnote{We note that for the special case of binary symbols, i.e., $\mathcal{S}=\{0,1\}$, the problem in (\ref{Eq:ML_NonCoherent}) can also be further simplified, cf. \cite{NanoCOM16}. However, since this simplification is not needed for further development in this paper, we do not pursue it here.}.  For future reference,  let $\omega(\mathbf{s})=\sum_{k=1}^K  s[k]$ denote the weight of sequence $\mathbf{s}$ and let $\omega_{\ell}(\mathbf{s},\mathbf{r})=\sum_{k=1}^K r[k] \mathbf{1}\{s[k]=\eta_{\ell}\}$ denote the weight of the observation sequence $\mathbf{r}$ corresponding to the positions where $s[k]=\eta_{\ell}$.

\begin{lem}\label{Lem:ML}
The ML sequence for coherent detection in (\ref{Eq:ML}) can be expressed as
\begin{align} \label{Eq:ML_Sol}
  \hat{\mathbf{s}}^{\mathrm{c}}  & = \underset{\mathbf{s}\in\boldsymbol{\mathcal{S}}}{\mathrm{argmax}} \,\,  
    \Lambda^{\mathrm{ML}}(\mathbf{s}),
\end{align}
where $\Lambda^{\mathrm{ML}}(\mathbf{s})= 
    - \omega(\mathbf{s}) \bar{c}_{\mathrm{s}}  + \sum_{{\ell}=1}^{L-1} \omega_{\ell}(\mathbf{s},\mathbf{r}) \mathsf{ln}\left( 1+ \eta_{\ell} \mathsf{SIR}  \right)$.
\end{lem}
\begin{IEEEproof}
The proof is provided in Appendix~\ref{App:LemML}.
\end{IEEEproof}

The following insights can be obtained from the optimal coherent ML solution in Lemma~\ref{Lem:ML}.

\begin{itemize}
\item Only variables $\omega(\mathbf{s})$ and $\omega_{\ell}(\mathbf{s},\mathbf{r})$, which are both functions of the hypothesis sequence, determine  the optimal ML decision. Hereby, $\omega(\mathbf{s})$ depends solely on the hypothesis sequence whereas $\omega_{\ell}(\mathbf{s},\mathbf{r})$ depends on both the hypothesis sequence and the observation~vector.
\item The variable $\omega_{\ell}(\mathbf{s},\mathbf{r})$ is multiplied by the weight $\mathsf{ln}\left( 1+ \eta_{\ell} \mathsf{SIR}  \right)$ which is a monotonically increasing function of $\eta_{\ell}$. Moreover, since by convention, we assumed $\eta_0=0$, $\mathsf{ln}\left( 1+ \eta_{\ell} \mathsf{SIR}  \right)=0$ holds. Therefore, weight $\omega_{0}(\mathbf{s},\mathbf{r})$, i.e., the sum of the observed molecules at positions where $s[k]=0$ holds, does not affect the ML metric $\Lambda^{\mathrm{ML}}(\mathbf{s})$ for sequence $\mathbf{s}$. For a binary symbol alphabet, i.e., $\mathcal{S}=\{0,1\}$, only  observations corresponding to the positions of ones in the hypothesis sequence affect the ML metric. 
\end{itemize} 
We employ the above insights in the next subsection to develop a CSI-free detection algorithm.

\subsection{CSI-Free Sequence Detection}

The definition of \textit{SCW codes} is formally presented in the following.
 
\begin{defin}\label{Def:StrConstCode}
Let SCW codes be denoted by $\boldsymbol{\mathcal{S}}^{\mathrm{sc}}(\bar{\boldsymbol{\omega}})$ with weight vector $\bar{\boldsymbol{\omega}}=[\bar{\omega}_0,\bar{\omega}_1,\dots, \bar{\omega}_{L-1}]^{\mathsf{T}}$. For an SCW code, all codewords $\mathbf{s}$ in the codebook meet the following condition  
\begin{IEEEeqnarray}{lll} \label{Eq:StrWeightCode}
\sum_{k=1}^{K}\mathbf{1}\{s[k]=\eta_{\ell}\} = \bar{\omega}_{\ell},\quad \forall \eta_{\ell}\in\mathcal{S} \,\,\text{and}\,\,\forall \mathbf{s} \in \boldsymbol{\mathcal{S}}^{\mathrm{sc}}(\bar{\boldsymbol{\omega}}). 
\end{IEEEeqnarray}
An SCW code is called a full code if all possible codewords that satisfy  (\ref{Eq:StrWeightCode}) are included in the codebook. Moreover, an SCW code is called balanced if all weights $\bar{\omega}_{\ell}$ are identical, i.e., $\bar{\omega}_{\ell}=\bar{\omega},\,\,\forall \ell$ holds. 
\hfill \QEDwhite
\end{defin}

\begin{remk}
CW codes, denoted by $\boldsymbol{\mathcal{S}}^{\mathrm{c}}(K,\omega)$, have been widely employed in conventional communication systems \cite{WeightCodeClass,CW_qray,BalancedCode_Knuth,MutiplyCW}. For these codes, weight $\omega(\mathbf{s})=\omega$ is constant for all codewords in the codebook. Obviously, an SWC code $\boldsymbol{\mathcal{S}}^{\mathrm{sc}}(\bar{\boldsymbol{\omega}})$ is also a CW code $\boldsymbol{\mathcal{S}}^{\mathrm{c}}(K,\omega)$ with $K=\sum_{\ell=0}^{L-1}\bar{\omega}_{\ell}$ and $\omega=\sum_{\ell=0}^{L-1}\bar{\omega}_{\ell}\eta_{\ell}$. We note that for binary codes, i.e., $\mathcal{S}=\{0,1\}$, CW codes and SCW codes become equivalent, i.e., $\boldsymbol{\mathcal{S}}^{\mathrm{sc}}([\bar{\omega}_0,\bar{\omega}_1]^{\mathsf{T}})=\boldsymbol{\mathcal{S}}^{\mathrm{c}}(K,\omega)$ where $\omega=\bar{\omega}_1=K-\bar{\omega}_0$.
\end{remk}

The following example illustrates several SCW codes and the corresponding CW codes. 
\begin{examp}
Let the length of the codewords be $K=6$. 
\begin{itemize}
\item First, we consider binary codes, i.e., $\mathcal{S}=\{0,1\}$. 
\begin{itemize}
\item $\mathbf{s}=[1,1,0,0,0,0]^{\mathsf{T}}$, $\mathbf{s}'=[0,1,0,0,0,1]^{\mathsf{T}}$, and $\mathbf{s}''=[0,0,1,1,0,0]^{\mathsf{T}}$ are example codewords of the SCW code $\boldsymbol{\mathcal{S}}^{\mathrm{sc}}\left([4,2]^{\mathsf{T}} \right)$  or equivalently the CW code $\boldsymbol{\mathcal{S}}^{\mathrm{c}}(6,2)$. These codes are equivalent to MPPM \cite{MPPM_Optic}.
\item $\mathbf{s}=[1,1,1,0,0,0]^{\mathsf{T}}$, $\mathbf{s}'=[0,1,1,0,0,1]^{\mathsf{T}}$, and $\mathbf{s}''=[0,0,0,1,1,1]^{\mathsf{T}}$ are example codewords of the balanced SCW code $\boldsymbol{\mathcal{S}}^{\mathrm{sc}}\left([3,3]^{\mathsf{T}} \right)$  or equivalently the balanced CW code $\boldsymbol{\mathcal{S}}^{\mathrm{c}}(6,3)$ \cite{BalancedCode_Knuth}.
\item $\mathbf{s}=[1,0,0,0,0,0]^{\mathsf{T}}$, $\mathbf{s}'=[0,1,0,0,0,0]^{\mathsf{T}}$, and $\mathbf{s}''=[0,0,0,0,1,0]^{\mathsf{T}}$ are example codewords of the SCW code $\boldsymbol{\mathcal{S}}^{\mathrm{sc}}\left([5,1]^{\mathsf{T}} \right)$  or equivalently the CW code $\boldsymbol{\mathcal{S}}^{\mathrm{c}}(6,1)$. These codes are equivalent to PPM \cite{PPM_Pieroborn}.
\end{itemize} 
\item Next, we consider ternary codes, e.g., $\mathcal{S}=\{0,0.5,1\}$.
\begin{itemize}
\item  $\mathbf{s}=[1,0.5,0.5,0,0,0]^{\mathsf{T}}$, $\mathbf{s}'=[0.5,1,0,0,0,0.5]^{\mathsf{T}}$, and $\mathbf{s}''=[0,0,0.5,0,0.5,1]^{\mathsf{T}}$ are example codewords of the SCW code $\boldsymbol{\mathcal{S}}^{\mathrm{sc}}\left([3,2,1]^{\mathsf{T}} \right)$  or equivalently the CW code $\boldsymbol{\mathcal{S}}^{\mathrm{c}}(6,2)$.
\item  $\mathbf{s}=[1,1,0.5,0.5,0,0]^{\mathsf{T}}$, $\mathbf{s}'=[0.5,1,0,0,1,0.5]^{\mathsf{T}}$, and $\mathbf{s}''=[1,0,0.5,0,0.5,1]^{\mathsf{T}}$ are example codewords of the balanced SCW code $\boldsymbol{\mathcal{S}}^{\mathrm{sc}}\left([2,2,2]^{\mathsf{T}} \right)$  or equivalently the CW code $\boldsymbol{\mathcal{S}}^{\mathrm{c}}(6,3)$.
\end{itemize}

\end{itemize}
\end{examp}

 The following theorem reveals how the ML sequence can be obtained without instantaneous or statistical CSI if a full SCW code is employed.

\begin{theo}\label{Theo:ML_StrCW}
Assuming a \textit{full} SCW code is employed, i.e., $\mathbf{s}\in\boldsymbol{\mathcal{S}}^{\mathrm{sc}}(\bar{\boldsymbol{\omega}})$, the solutions of (\ref{Eq:ML}) and (\ref{Eq:ML_NonCoherent}) are identical and independent of both  \textit{instantaneous} CSI ($\bar{c}_{\mathrm{s}}$ and $\bar{c}_{\mathrm{n}}$) and  \textit{statistical} CSI ($f_{\bar{\mathbf{c}}}(\bar{c}_{\mathrm{s}},\bar{c}_{\mathrm{n}})$). This enables optimal CSI-free detection based on Algorithm~\ref{Alg:ML_StrCW}. Moreover, for a full \textit{binary} CW code, $\boldsymbol{\mathcal{S}}^{\mathrm{c}}(K,\omega)$,  the solution of (\ref{Eq:ML}) and (\ref{Eq:ML_NonCoherent}) is simply the codeword whose ``1" elements  correspond to the $\omega$ largest elements of $\mathbf{r}$. 
\end{theo}
\begin{IEEEproof}
The proof is provided in Appendix~\ref{App:ML_StrCW}.
\end{IEEEproof}

\begin{algorithm}[t] \label{Alg:ML_StrCW}
\caption{ML Sequence Detection for SCW Codes}
\begin{algorithmic}[1] 
\STATE \textbf{initialize} Sort observation vector $\mathbf{r}$ in ascending order into a new vector $\tilde{\mathbf{r}}$. 
\STATE Set those elements of $\mathbf{s}$ which correspond to the $\bar{\omega}_{0}$ first elements of $\tilde{\mathbf{r}}$ to $\eta_0=0$.
\FOR{$\ell=1$ until $\ell=L-1$}
        \STATE Set those elements of $\mathbf{s}$ which correspond to element $\sum_{\ell'=0}^{\ell-1}\bar{\omega}_{\ell'}+1$ to element $\sum_{\ell'=0}^{\ell-1}\bar{\omega}_{\ell'}+\bar{\omega}_{\ell}$ of $\tilde{\mathbf{r}}$ to $\eta_{\ell}$.
 \ENDFOR
 \STATE Return $\mathbf{s}$ as the ML sequence.
\end{algorithmic}
\end{algorithm}

We note that the ML sequence is not necessarily unique, i.e., more than one sequence may achieve the maximum value of the likelihood function in (\ref{Eq:ML}) and (\ref{Eq:ML_NonCoherent}). This can be also seen from Algorithm~\ref{Alg:ML_StrCW} where the ordered vector $\tilde{\mathbf{r}}$ may not necessarily be unique since some elements of $\mathbf{r}$ can be identical.   To further explain the optimal sequence detector for SCW codes in Algorithm~\ref{Alg:ML_StrCW}, we present the following examples.

\begin{examp} Suppose an SCW code with symbol set $\mathcal{S}=\{0,0.5,1\}$ and weight vector $\bar{\boldsymbol{\omega}}=[2, 3, 1]^{\mathsf{T}}$ is employed and we wish to decode the observation vector $\mathbf{r}=[12, 4, 8, 6, 15, 10]^{\mathsf{T}}$. 

\begin{itemize}
\item In line 1 of Algorithm~\ref{Alg:ML_StrCW}, $\mathbf{r}$ is reordered in ascending order into vector $\tilde{\mathbf{r}}=[4, 6, 8, 10, 12, 15]^{\mathsf{T}}$.
\item In line 2 of Algorithm~\ref{Alg:ML_StrCW}, the two elements ($\bar{\omega}_{0}=2$) of $\mathbf{s}$ corresponding to the first two elements of $\tilde{\mathbf{r}}$ are set to $\eta_0=0$.  This leads to $\mathbf{s}= [\times,0,\times,0,\times,\times]^{\mathsf{T}}$.
\item In line 4 of Algorithm~\ref{Alg:ML_StrCW}, the three elements ($\bar{\omega}_{1}=3$) of $\mathbf{s}$ corresponding to the third to the fifth elements of $\tilde{\mathbf{r}}$ are set to $\eta_1=0.5$.  This leads to $\mathbf{s}= [0.5,0,0.5,0,\times,0.5]^{\mathsf{T}}$.
\item In line 4 of Algorithm~\ref{Alg:ML_StrCW}, the one remaining element ($\bar{\omega}_{2}=1$) of $\mathbf{s}$ corresponding to the sixth element of $\tilde{\mathbf{r}}$ is set to $\eta_2=1$. This leads to the ML sequence $\mathbf{s}= [0.5,0,0.5,0,1,0.5]^{\mathsf{T}}$ which is returned in line 6 of Algorithm~\ref{Alg:ML_StrCW}.
\end{itemize}   
\end{examp} 

\begin{examp} Suppose a balanced binary CW code of length $K=6$, i.e., $\mathcal{S}=\{0,1\}$ and $\omega=3$, is employed and we wish to decode the observation vector $\mathbf{r}=[12, 4, 8, 6, 15, 8]^{\mathsf{T}}$. According to Theorem~\ref{Theo:ML_StrCW}, the optimal sequence is  the codeword whose ``1" elements  correspond to the $\omega=3$ largest elements of $\mathbf{r}$, i.e., elements $15$, $12$, and $8$. However, since we have two elements with value $8$, we obtain two ML sequences as $\mathbf{s}= [1,0,0,0,1,1]^{\mathsf{T}}$ and $\mathbf{s}=[1,0,1,0,1,0]^{\mathsf{T}}$ of which one has to be picked at random.  
\end{examp} 

\begin{remk}
  We note that the length of observation vector $\mathbf{r}$, which needs to be sorted into $\tilde{\mathbf{r}}$, and the number of assignment operations in each iteration of the for-loop in Algorithm~\ref{Alg:ML_StrCW}, \textit{proportionally} increase with the codeword length $K$. Therefore, the complexity of Algorithm~\ref{Alg:ML_StrCW} is linear in the codeword length, $K$. Moreover, asymptotically for large $K$,  the sorting operation can be performed with a complexity on the order of $O(K\mathsf{log}(\mathsf{log}(L)))$ according to the \textit{Van Emde Boas tree} \cite{SortingComplexity}. Note that for the general coherent and non-coherent ML problems in (\ref{Eq:ML}) and (\ref{Eq:ML_NonCoherent}), the complexity is exponential in $K$ since the number of codewords and hence, the number of metrics which need to be computed, grow exponentially in $K$. Therefore, the proposed SCW codes do not only avoid the complexity and challenges associated with CSI acquisition but also significantly reduce the complexity of ML detection. This makes SCW codes particularly suitable for simple nano-machines with limited computational capabilities.
\end{remk}

\begin{remk}
We emphasize that CSI-free detection of SCW codes is possible provided that the adopted codebook is full. However, the number of possible SCW codewords is usually not a power of two which complicates the bit-to-codeword (bit-to-symbol) mapping. In particular, to fully exploit all possible codewords, one has to perform a multi-dimensional bit-to-codeword mapping. One straightforward approach to obtain a simple bit-to-codeword mapping is to map some of the bit sequences to more than one codeword. In this way, at the cost of decreasing the code rate, the full codebook is employed and CSI-free detection with Algorithm~\ref{Alg:ML_StrCW} is still applicable.  Alternatively, one may employ a subset of all possible codewords, use the detector in Algorithm~\ref{Alg:ML_StrCW}, and declare a decoding error if a codeword, which does not belong to the adopted codebook, is detected.   
\end{remk}

While Theorem~\ref{Theo:ML_StrCW} claims CSI-free detection for full SCW codes,  in the following, we show that for binary CW codes, CSI-free detection is possible even if the codebook is not full.  

\begin{corol}\label{Corol:BinaryCW}
For \textit{binary} CW codes (not necessarily full codes), i.e., $\mathbf{s}\in\boldsymbol{\mathcal{S}}^{\mathrm{c}}(K,\omega)$ and $\mathcal{S}=\{0,1\}$, the solutions of (\ref{Eq:ML}) and (\ref{Eq:ML_NonCoherent}) are identical and require neither instantaneous CSI  nor statistical CSI. In this case, the optimal CSI-free decision is obtained from
\begin{align} \label{Eq:ML_BinSol}
  \hat{\mathbf{s}} &= \underset{\mathbf{s}\in\boldsymbol{\mathcal{S}}^{\mathrm{c}}(K,\omega)}{\mathrm{argmax}} \,\, \omega_{1}(\mathbf{s},\mathbf{r}) = \underset{\mathbf{s}\in\boldsymbol{\mathcal{S}}^{\mathrm{c}}(K,\omega)}{\mathrm{argmax}} \,\, \sum_{k=1}^K s[k]r[k].
\end{align}
\end{corol}
 \begin{IEEEproof}
The proof  follows directly from substituting binary symbols, i.e., $\mathcal{S}=\{0,1\}$, into (\ref{Eq:ML_Sol}) in Lemma~\ref{Lem:ML}.
 \end{IEEEproof}

\section{Performance Analysis}
In this section, we analyze the code rate, error rate, and average number of released molecules for the proposed SCW codes.

 \subsection{Rate Analysis}
 
The rate of a general code comprised of $M$ codewords of length $K$ with symbol set $\mathcal{S}$ is given by 
\begin{IEEEeqnarray}{lll} \label{Eq:CodeRateGen}
 R^{\mathrm{code}}(\bar{\boldsymbol{\omega}}) = \frac{\mathsf{log}\left(M\right)}{\mathsf{log}\left(|\mathcal{S}|^K\right)} = \frac{\mathsf{log}_{|\mathcal{S}|}\left(M\right)}{K}.
\end{IEEEeqnarray}
We note that the code rate specifies the information content of a codeword compared to uncoded transmission with the same symbol set. Therefore, the code rate in (\ref{Eq:CodeRateGen}) is unitless. Alternatively, one can define the information rate or data rate in bits/symbol, denoted by $R^{\mathrm{inf}}(\bar{\boldsymbol{\omega}})$, as the average number of information bits that a symbol in a codeword contains. The relation between $R^{\mathrm{code}}(\bar{\boldsymbol{\omega}})$ and $R^{\mathrm{inf}}(\bar{\boldsymbol{\omega}})$ is given by
\begin{IEEEeqnarray}{lll} \label{Eq:InfRate}
 R^{\mathrm{inf}}(\bar{\boldsymbol{\omega}}) = \frac{\mathsf{log}_{2}\left(M\right)}{K}
 =\frac{1}{\mathsf{log}_2(L)}R^{\mathrm{code}}(\bar{\boldsymbol{\omega}}) \quad\text{bits/symbol}. \quad\,\,
\end{IEEEeqnarray}
The code rate of a full SCW code is an upper bound for the code rate of SCW codes that do not use all possible codewords. Hence, in the following, we consider the code rate of full SCW~codes.
 
 \begin{prop}\label{Prop:CodeRate}
 The code rate of a full SCW code, $\boldsymbol{\mathcal{S}}^{\mathrm{sc}}(\bar{\boldsymbol{\omega}})$, is given by
\begin{IEEEeqnarray}{lll} \label{Eq:StrCT_Rate}
R^{\mathrm{code}}(\bar{\boldsymbol{\omega}})&=\frac{1}{\sum_{\ell=1}^K \bar{\omega}_{\ell}}\sum_{\ell =0}^{L-1} \mathsf{log}_{L}\left({\sum_{\ell'\leq\ell} \bar{\omega}_{\ell'}\choose \bar{\omega}_{\ell}}\right) \nonumber \\
&=  \frac{1}{K}\mathsf{log}_{L}\left(\frac{K!}{\prod_{\ell=0}^{L-1}\bar{\omega}_{\ell}!}\right) \overset{K\to\infty}{\to}  H_L(\boldsymbol{\rho}),
\end{IEEEeqnarray}
where $\boldsymbol{\rho}=[\rho_0,\rho_1,\dots,\rho_L]^{\mathsf{T}}$ and $\rho_{\ell}=\bar{\omega}_{\ell}/K$.
 \end{prop}
 \begin{IEEEproof}
The proof is provided in Appendix~\ref{App:PropRate}.
 \end{IEEEproof}

Given $K$ and $L$, the code rate of SCW codes is maximized when they are balanced, i.e., $\bar{\omega}_{\ell}=\bar{\omega}_{\ell'},\,\,\forall \ell,\ell'$ assuming $K/L$ is an integer. Moreover, for balanced codes, the rate approaches $R^{\mathrm{code}}(\bar{\boldsymbol{\omega}})\to 1$ as $K\to\infty$.  In the following, we provide simple upper and lower bounds for the special case of full \textit{binary} CW codes.

 \begin{corol}\label{Corol:RateBin}
 There exists an $\alpha\in[\sqrt{2\pi}/e^2,e/2\pi]$ such that the following equation holds for the code rate of a full binary CW code, $\boldsymbol{\mathcal{S}}^{\mathrm{c}}(K,\omega)$: 
\begin{IEEEeqnarray}{lll} \label{Eq:CodeRateBin}
R^{\mathrm{code}}(K,\omega)  = H_2(\rho)-\frac{1}{K}\mathsf{log}_2 \left(\frac{\sqrt{\rho(1-\rho)K}}{\alpha}\right)
\overset{K\to\infty}{\to}H_2(\rho),
\end{IEEEeqnarray}
where $\rho=\omega/K$. In other words, substituting the lower and upper limits of interval $[\sqrt{2\pi}/e^2,e/2\pi]\approx[0.3392,0.4326]$ for $\alpha$ in (\ref{Eq:CodeRateBin}) yields lower and upper bounds on the~code~rate. 
 \end{corol}
 \begin{IEEEproof}
The proof is provided in Appendix~\ref{App:CorolRateBin}.
 \end{IEEEproof}
 
Note that  for even values of $K$, the bounds on the code rate of the corresponding balanced binary code simplify to $R(K,K/2)=1-\mathsf{log}_2(\sqrt{K}/2\alpha)/K$.

\subsection{Error Analysis}
Let $P_e^{\mathrm{code}}(\bar{\boldsymbol{\omega}}|\bar{\mathbf{c}})$ denote the codeword error rate (CER) of the SCW code with weight $\bar{\boldsymbol{\omega}}$ for a given realization of the CSI $\bar{\mathbf{c}}$. In the following, we provide several analytical bounds for the CER $P_e^{\mathrm{code}}(\bar{\boldsymbol{\omega}}|\bar{\mathbf{c}})$.  First, we present an upper bound on the CER based on the pairwise error probability (PEP) and union and Chernoff bounds.

\begin{prop}\label{Prop:UpperGen}
The CER of the optimal detector for SCW codes, $\boldsymbol{\mathcal{S}}^{\mathrm{sc}}(\bar{\boldsymbol{\omega}})$,  is upper bounded~by
\begin{IEEEeqnarray}{lll} \label{Eq:UpperGen}
P_e^{\mathrm{code}}(\bar{\boldsymbol{\omega}}|\bar{\mathbf{c}}) 
 &\leq  \frac{1}{M} \sum_{\forall\mathbf{s}} \sum_{\forall\hat{\mathbf{s}}\neq \mathbf{s}} \mathsf{exp}\left(\sum_{k=1}^K \lambda[k]\left(\mathsf{exp}\left(\varpi[k]t\right)-1\right)\right),  \nonumber \\
 &\overset{(a)}{=}\sum_{\forall\hat{\mathbf{s}}\neq \mathbf{s}} \mathsf{exp}\left(\sum_{k=1}^K \lambda[k]\left(\mathsf{exp}\left(\varpi[k]t\right)-1\right)\right),\,\,\forall t>0, 
\end{IEEEeqnarray}
where $\lambda[k]=s[k]\bar{c}_{\mathrm{s}}+\bar{c}_{\mathrm{n}}$ and $\varpi[k]=\mathsf{ln}\left( \frac{1+ \hat{s}[k] \mathsf{SIR}} {1+ s[k] \mathsf{SIR}} \right)$. Moreover, equality $(a)$ holds only if the adopted SCW code is full. Thereby, for equality $(a)$, $\mathbf{s}$ can be any arbitrary  codeword chosen from the codebook. In (\ref{Eq:UpperGen}), $t$ is an arbitrary positive real number which is introduced by the Chernoff bound that was used to arrive at the upper bound.
\end{prop}
 \begin{IEEEproof}
The proof is provided in Appendix~\ref{App:PropUpperGen}.
 \end{IEEEproof}

 We note that (\ref{Eq:UpperGen}) constitutes an upper bound on the CER for any value of $t>0$. Therefore, one can optimize $t$ to tighten the upper bound.    For notational simplicity,  we enumerate the codewords by $\mathbf{s}_i,\,\,i=1,\dots,M$. Moreover, let $d_{ij}=h(\mathbf{s}_i,\mathbf{s}_j)$ be the Hamming distance between codewords $\mathbf{s}_i$ and $\mathbf{s}_j$. In the following corollary, we present a tighter upper bound than the general upper bound presented in Proposition~\ref{Prop:UpperGen} for binary~CW~codes. 

\begin{corol}\label{Corol:CER_UppBin}
The CER of the optimal detector for binary CW code, $\boldsymbol{\mathcal{S}}^{\mathrm{c}}(K,\omega)$,  is upper bounded~by
\begin{IEEEeqnarray}{lll} 
P_e^{\mathrm{code}}(K,\omega |\bar{\mathbf{c}}) \leq \frac{1}{M} \sum_{\forall d_{ij},\,i\neq j} 0.5f_X(0)+\sum_{x=1}^{\infty}
 f_X(x),\quad\,\, \label{Eq:CER_UppBin}
\end{IEEEeqnarray}
where $f_X(x)$ is given by 
\begin{IEEEeqnarray}{lll} 
 f_X(x) = e^{-(\lambda_1+\lambda_2)}\left(\frac{\lambda_2}{\lambda_1}\right)^{x/2}I_x(2\sqrt{\lambda_1\lambda_2}), \label{Eq:Skellam}
\end{IEEEeqnarray}
with $\lambda_1=\frac{d_{ij}(\bar{c}_{\mathrm{s}}+\bar{c}_{\mathrm{n}})}{2}$, $\lambda_2=\frac{d_{ij}\bar{c}_{\mathrm{n}}}{2}$, and $I_x(\cdot)$ is the modified Bessel function of the first kind and order $x$ \cite{TableIntegSerie}.
\end{corol}
 \begin{IEEEproof}
The proof is provided in Appendix~\ref{App:Corol_CER_UppBin}.
\end{IEEEproof}

The upper bounds in Proposition~\ref{Prop:UpperGen} and Corollary~\ref{Corol:CER_UppBin} are based on the PEP and the union bound. Hence, they are expected to be tight at high SINRs. In the following proposition, we provide upper and lower bounds on the CER for the special case of \textit{full} binary CW codes which are tight for all SINRs.

\begin{prop}\label{Prop:UppBinFull}
The CER of the optimal detector for a full binary CW code, $\boldsymbol{\mathcal{S}}^{\mathrm{c}}(K,\omega)$,~is~bounded~as
\begin{IEEEeqnarray}{lll} 
\Scale[1]{\displaystyle \sum_{y=1}^{\infty} \hspace{-0.5mm} F_X(y-1)f_Y(y) 
\leq P_e^{\mathrm{code}}(K,\omega |\bar{\mathbf{c}})  
\leq \sum_{y=0}^{\infty} \hspace{-0.5mm} F_X(y)f_Y(y),\quad\,\,} \,\, \label{Eq:CER_sum}
\end{IEEEeqnarray}
where $F_X(\cdot)$ and $f_Y(\cdot)$ are given by
\begin{IEEEeqnarray}{rll} \label{Eq:PDFMaxMin}
F_X(x) &\,\,= 1- (1-F_{\mathcal{P}}(x,\bar{c}_{\mathrm{s}}+\bar{c}_{\mathrm{n}}))^{\omega} \qquad \IEEEyesnumber \IEEEyessubnumber \\
f_Y(y) &\,\,= (K-\omega) f_\mathcal{P}(y,\bar{c}_{\mathrm{n}}) F_{\mathcal{P}}(y,\bar{c}_{\mathrm{n}})^{K-\omega -1}.  \IEEEyessubnumber 
\end{IEEEeqnarray}
In (\ref{Eq:PDFMaxMin}a) and (\ref{Eq:PDFMaxMin}b), $f_\mathcal{P}(\cdot,\cdot)$ and $F_{\mathcal{P}}(\cdot,\cdot)$ are given by
\begin{IEEEeqnarray}{rll} \label{Eq:PDFPoisson}
f_\mathcal{P}(x,\lambda) &= \frac{\lambda^{x}e^{-\lambda}}{x!} \qquad \IEEEyesnumber \IEEEyessubnumber \\
F_\mathcal{P}(x,\lambda) &= Q(\lfloor x+1\rfloor,\lambda),  \IEEEyessubnumber 
\end{IEEEeqnarray}
where $Q(\cdot,\cdot)$ is the regularized Gamma function \cite{TableIntegSerie}.
\end{prop}
 \begin{IEEEproof}
The proof is provided in Appendix~\ref{App:PropUppBinFull}.
\end{IEEEproof}
  
    \begin{remk}
  In Propositions~\ref{Prop:UpperGen} and \ref{Prop:UppBinFull} and Corollary~\ref{Corol:CER_UppBin}, we proposed different bounds on the CER of SCW codes.   We note that, for any code, the relation between  the bit error rate (BER), denoted by $P_e^{\mathrm{bit}}(\bar{\boldsymbol{\omega}}|\bar{\mathbf{c}})$, and the CER depends on mapping being used to map the bit sequences to the codewords (hyper-symbols). However, in general, $P_e^{\mathrm{bit}}(\bar{\boldsymbol{\omega}})< \mu P_e^{\mathrm{code}}(\bar{\boldsymbol{\omega}}|\bar{\mathbf{c}})$, where $\mu\in[0,1]$, holds for the  relation between CER and BER. For instance, for binary PPM with $K=2$, i.e.,  $\boldsymbol{\mathcal{S}}^{\mathrm{c}}(2,1)$ with codewords $\mathbf{s}\in\{[1,0]^{\mathsf{T}},[0,1]^{\mathsf{T}}\}$, we have $P_e^{\mathrm{bit}}(\bar{\boldsymbol{\omega}}|\bar{\mathbf{c}})=P_e^{\mathrm{code}}(\bar{\boldsymbol{\omega}}|\bar{\mathbf{c}})$, i.e., $\mu=1$. On the other hand, assuming large $K$ and random mapping\footnote{Here, by random mapping, we mean a mapping strategy where the bit sequence-codeword pairs are formed at random. Nevertheless, for online transmission, the mapping is fixed and known to transmitter and receiver, of course.} of bit sequences to codewords, we obtain $P_e^{\mathrm{bit}}(\bar{\boldsymbol{\omega}}|\bar{\mathbf{c}})=0.5 P_e^{\mathrm{code}}(\bar{\boldsymbol{\omega}}|\bar{\mathbf{c}})$, i.e., in case of a codeword error, the original bit sequence is decoded as a different random bit sequence. An interesting related research problem is the design of mapping strategies which minimize the BER. In fact, one may design a mapping strategy that ensures the codewords with the highest pairwise error probability are mapped to bit sequences which have the minimum possible Hamming distance. In this paper, we do not investigate this problem  due to space constraints and leave it for future~work.
  \end{remk}

\subsection{Energy Analysis}

In Section~II, we assumed that the maximum number of molecules that the transmitter can release in one symbol interval is limited to $N^{\mathrm{tx}}$, i.e., a peak power constraint is adopted. Using CW codes  implies that the number of  molecules released by the transmitter of the considered MC system is equal to $N^{\mathrm{tx}}\omega$ for all codewords. Therefore, the average number of  molecules released per symbol interval, denoted by $\bar{N}^{\mathrm{tx}}(\bar{\boldsymbol{\omega}})$,  is given by $\bar{N}^{\mathrm{tx}}(\bar{\boldsymbol{\omega}})=\frac{\omega}{K}N^{\mathrm{tx}}$. Hence, for SCW code $\boldsymbol{\mathcal{S}}^{\mathrm{sc}}(\bar{\boldsymbol{\omega}})$, the average number of molecules released by the transmitter  is obtained as
\begin{IEEEeqnarray}{rll} \label{Eq:AvgMolec}
\bar{N}^{\mathrm{tx}}(\bar{\boldsymbol{\omega}}) = \frac{\sum_{\ell=0}^{L-1}\bar{\omega}_{\ell}\eta_{\ell}}{\sum_{\ell=0}^{L-1}\bar{\omega}_{\ell}} N^{\mathrm{tx}}.
\end{IEEEeqnarray}

When evaluating the performance of SCW codes for MCs, one may consider the trade-off between rate performance, error performance, and average energy consumption. In particular, the rate performance quantifies how fast the information bits can be transmitted, the error performance is a measure for the reliability of communication, and the energy consumption is related to the average number of  molecules released by the transmitter.

\subsection{Balanced Codes}

To gain further insight, let us focus on balanced codes, i.e., $\bar{\omega}_{\ell}=\frac{K}{L},\,\,\forall\ell$, assuming $K/L$ is an integer, and the following symbol set
\begin{align} \label{Eq:SymbSet}
  \mathcal{S}=\left\{0,\frac{1}{L-1},\frac{2}{L-1},\cdots,\frac{L-2}{L-1},1\right\}.
\end{align}
Some notable results for the above special case are provided in the following.

\subsubsection{Rate Performance}

We first note that the code rate in (\ref{Eq:StrCT_Rate}) depends on the number of symbols, $L$, but is not a function of the symbol set, $\mathcal{S}$. Substituting weights $\bar{\omega}_{\ell}=\frac{K}{L},\,\,\forall\ell$, into (\ref{Eq:StrCT_Rate}), we obtain
\begin{IEEEeqnarray}{lll} \label{Eq:RateSpecCase}
R^{\mathrm{code}}(\bar{\boldsymbol{\omega}})&=  \frac{1}{K}\mathsf{log}_{L}\left(\frac{K!}{\left(\frac{K}{L}!\right)^L}\right) 
= \frac{1}{K}\left[\mathsf{log}_{L}\left(K!\right) -L\mathsf{log}_{L}\left(\frac{K}{L}!\right)\right].
\end{IEEEeqnarray}
We note that the code rate in (\ref{Eq:RateSpecCase}) is a decreasing function of $L$; however, the proof of this property seems very involved. In Section~V, we will show that the code rate, $R^{\mathrm{code}}(\bar{\boldsymbol{\omega}})$, monotonically decreases in $L$ for several examples. Nevertheless, we emphasize that the data rate, $R^{\mathrm{inf}}(\bar{\boldsymbol{\omega}})$, is still a monotonically increasing function of $L$.

\subsubsection{Error Performance}

Since the CER does not lend itself to a simple expression  even for the special case considered here, we study the minimum distance between the codewords as a measure for reliability. In particular, for full balanced SCW codes with the symbol set in (\ref{Eq:SymbSet}), the minimum Euclidean distance, denoted by $d^{\mathsf{min}}$,  is obtained as
\begin{IEEEeqnarray}{rll} \label{Eq:MinDis}
d^{\mathsf{min}}  =\underset{\forall \mathbf{s}_i,\mathbf{s}_j\in\boldsymbol{\mathcal{S}}^{\mathrm{sc}}(\bar{\boldsymbol{\omega}})|i\neq j}{\mathsf{min}}\,\, \sqrt{\sum_{k=1}^{K}|s_i[k]-s_j[k]|^2} \,= \frac{\sqrt{2}}{L-1}. 
\end{IEEEeqnarray}
Note that any two codewords of an SCW code differ in at least two elements. Moreover, the minimum distance between two elements of an  SCW code with  the symbol set in (\ref{Eq:SymbSet}) is $\frac{1}{L-1}$. In fact, for a given $K$, the minimum distance in (\ref{Eq:MinDis}) decreases as $L$ increases which increases the CER. 

\subsubsection{Energy Performance}
The average number of released molecules for the balanced SCW code with the symbol set in (\ref{Eq:SymbSet}) can be simplified to
\begin{IEEEeqnarray}{rll} 
\bar{N}^{\mathrm{tx}}(\bar{\boldsymbol{\omega}}) = \frac{\frac{K}{L}\times\sum_{\ell=0}^{L-1}\frac{\ell}{L-1}}{L\times\frac{K}{L}} N^{\mathrm{tx}}
= \frac{1}{L(L-1)}\times\frac{(L-1)L}{2}N^{\mathrm{tx}}=\frac{1}{2}N^{\mathrm{tx}}.
\end{IEEEeqnarray}
Interestingly, the average number of released molecules for the balanced code with the symbol set in (\ref{Eq:SymbSet}) is not a function of the cardinality of the symbol set, $L$, and is constant, i.e., $0.5N^{\mathrm{tx}}$. We note that for uncoded transmission with equiprobable symbols taken from the symbol set in (\ref{Eq:SymbSet}), the average number of released molecules is also $0.5N^{\mathrm{tx}}$.

\section{Numerical Results}

In this section, we first discuss the simulation setup, i.e., the considered MC channel model and the adopted system parameters. Subsequently, we evaluate the performances of the proposed CSI-free~detector. 

\subsection{Simulation Setup}

Since the proposed detection scheme does not require CSI, it can be adopted regardless of whether the
channel is deterministic/time-invariant or stochastic/time-variant\footnote{We assume that the channel is fixed during one codeword. Therefore, if the MC channel is time-variant, the CSI may change only from one codeword to the next.}. In Figs.~\ref{Fig:CER_SCW}, \ref{Fig:CER_SCW_Bal}, \ref{Fig:CER_SNR}, and \ref{Fig:Tradeoff} b), we adopt the deterministic channel  with flow introduced in \cite{Adam_Enzyme}, and in Fig.~\ref{Fig:BER_SNR}, we consider the stochastic channel in \cite{NanoCOM16}.   In particular, both channel models  are based on the following equation for the \textit{expected} number of molecules observed at the receiver as a function of time 
\begin{align} \label{Eq:Consentration}
\hspace{-0.25cm}  \bar{c}_{\mathrm{s}}(t) = \frac{N^{\mathrm{tx}}V^{\mathrm{rx}}}{(4\pi D t)^{3/2}} \mathrm{exp}\left(-\kappa\bar{c}_{\mathrm{e}}t-\frac{(d-v_{\parallel}t)^2+(v_{\perp}t)^2}{4Dt}\right), \hspace{-0.15cm}
\end{align}
where the definition of the involved variables and their default values are provided in Table~I, see \cite{NanoCOM16,Adam_Enzyme} for detailed descriptions. We assume a symbol duration of $T^{\mathrm{symb}}=1$ ms and the receiver counts the number of molecules within its volume at sampling time $T^{\mathrm{samp}}=0.1$ ms after the beginning of a symbol interval. For instance, for the default values of the system parameters given in Table~I, we obtain $\bar{c}_{\mathrm{s}}= \bar{c}_{\mathrm{s}}(t=T^{\mathrm{samp}})=4.9$ molecules. Note that, assuming a fixed $\bar{c}_{\mathrm{n}}$, one may change the number of released molecules, $N^{\mathrm{tx}}$, to obtain different SINRs according to $\mathsf{SINR} =\frac{\bar{c}_{\mathrm{s}}^2}{\bar{c}_{\mathrm{s}}+\bar{c}_{\mathrm{n}}}$. Here, we assume $\bar{c}_{\mathrm{n}}=4.9$ which yields $\mathsf{SINR}\approx 4$ dB for the default values of the system parameters in Table~I. Finally, for the simulation results provided in this section, we choose the symbol set in (\ref{Eq:SymbSet}).

\begin{remk}
For the results presented in this section, we employ both full codebooks and partial codebooks which have a specific code rate.  To generate the partial codebook, we randomly select a given number of codewords from the full codebook. Which codewords are selected does not affect the code rate, but may significantly impact the error rate. Therefore, one may select the codewords such that the error rate is minimized. Since this is a challenging problem in general, one common approach is to select the codewords such that the average or minimum  distance between the selected codewords is maximized \cite{CodeMinDis}. However, for simplicity and to avoid the impact of specific codebook designs, we pick codewords at random to construct partial codebooks in this paper.
\end{remk}

\begin{table}
\label{Table:Parameter}
\caption{Default Values of the System Parameters \cite{NanoCOM16,Adam_Enzyme}. \vspace{-0.2cm}} 
\begin{center}
\scalebox{0.7}
{
\begin{tabular}{|| c | c | c ||}
  \hline 
  Variable & Definition & Value \\ \hline \hline
       $N^{\mathrm{tx}}$ & Number of released molecules  & $10^4$ molecules \\ \hline      
       $V^{\mathrm{rx}}$ & Receiver volume   & $\frac{4}{3}\pi 50^3$ \,\, ${\text{nm}}^3$ \\  
         &    & (a sphere with radius $50$ nm) \\ \hline   
        $d$ &  Distance between  transmitter and  receiver  & $500$ nm\\ \hline 
         $D$ &  Diffusion coefficient for the signaling molecule & $4.3\times 10^{-10}$ $\text{m}^2\cdot\text{s}^{-1}$\\ \hline          
       $\bar{c}_{\mathrm{e}}$ &  Enzyme concentration  & $10^{5}$ $\text{molecule}\cdot\mu\text{m}^3$ \\   
         &     & (approx. $1.66$ micromolar) \\ \hline  
       $\kappa$ &    Rate of molecule degradation reaction & $2\times10^{-19}$ $\text{m}^3\cdot\text{molecule}^{-1}\cdot\text{s}^{-1}$ \\ \hline 
       $(v_{\parallel},v_{\perp})$ &  Components of flow velocity   & $(10^{-3},10^{-3})$ $\text{m}\cdot\text{s}^{-1}$ \\ \hline
\end{tabular}
}
\end{center}\vspace{-0.5cm}
\end{table}

\subsection{Performance Evaluation}
In the following, we first verify the rate and error performance analyses provided in Propositions~\ref{Prop:CodeRate}, \ref{Prop:UpperGen}, and \ref{Prop:UppBinFull}, and Corollaries~\ref{Corol:RateBin} and \ref{Corol:CER_UppBin}. Subsequently, we illustrate the trade-off between rate, error performance, and average number of released molecule for an example and also compare the proposed CSI-free detector with some benchmark schemes  from the literature.

\subsubsection{Rate Analysis} 

\begin{figure}[!tbp]
  \centering
  \begin{minipage}[b]{0.49\textwidth}
  \centering
\resizebox{1\linewidth}{!}{\psfragfig{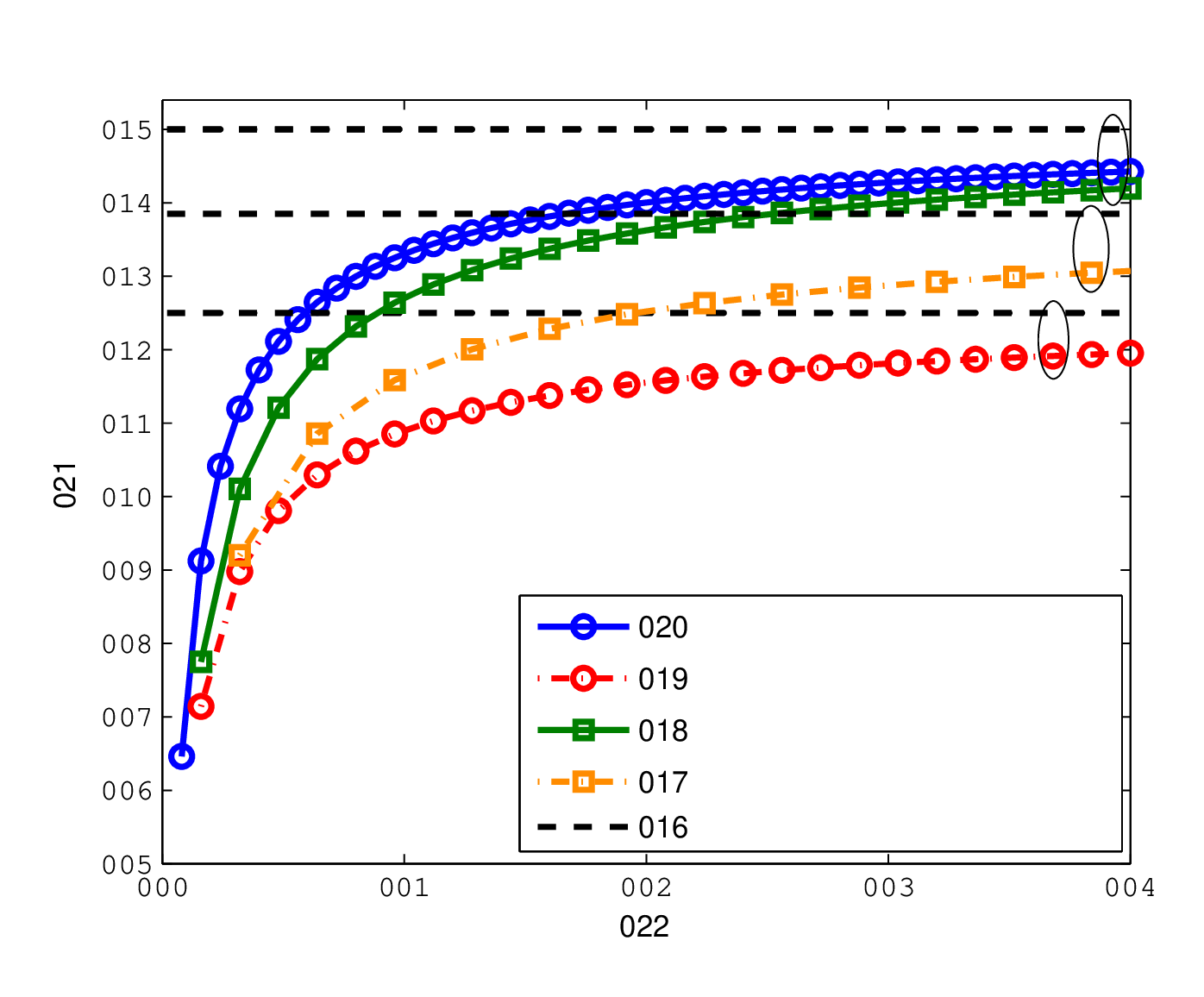}} \vspace{-0.8cm}
\caption{Code rate, $R^{\mathrm{code}}(\bar{\boldsymbol{\omega}})$, versus codeword length, $K$, for 
 balanced and unbalanced SWC codes with $L=4,8$. The curves with identical asymptotic code rate for $K\to\infty$  are labeled with an ellipse.
\vspace{-0.3cm} }
\label{Fig:RateKmulti}
  \end{minipage}
    \hfill
  \begin{minipage}[b]{0.01\textwidth}
  \end{minipage}
  \hfill
  \begin{minipage}[b]{0.49\textwidth}
  \centering
\resizebox{1\linewidth}{!}{\psfragfig{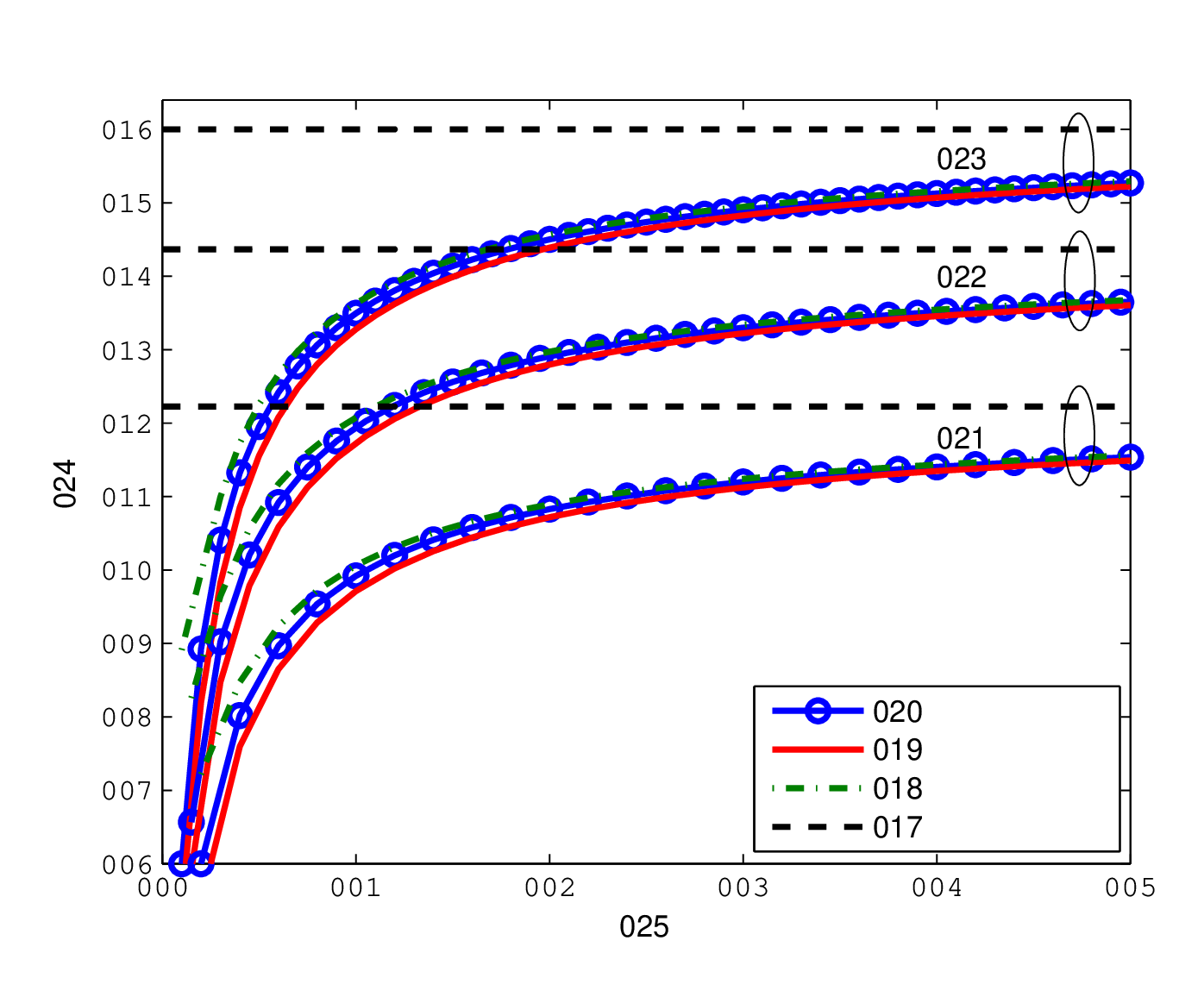}} \vspace{-0.8cm}
\caption{Code rate for binary CW codes, $R^{\mathrm{code}}(K,\omega)$, versus codeword length, $K$, for different $\rho \in\{\frac{1}{2},\frac{1}{3},\frac{1}{4}\}$. The curves with identical asymptotic code rate for $K\to\infty$  are labeled with an ellipse.\vspace{-0.3cm} }
\label{Fig:RateK}
  \end{minipage}
    \hfill
  \begin{minipage}[b]{0.01\textwidth}
  \end{minipage}\vspace{-0.4cm}
\end{figure}

First, using Proposition~\ref{Prop:CodeRate} and Corollary~\ref{Corol:RateBin}, we present some results for the code rate of the proposed SCW codes. In particular, in Fig.~\ref{Fig:RateKmulti}, the code rate $R^{\mathrm{code}}(\bar{\boldsymbol{\omega}})$, versus the codeword length, $K$, is shown for different cardinalities of the symbol set, i.e., $L=4,8$, and different code weights, i.e., $\bar{\boldsymbol{\omega}}$ or equivalently  $\boldsymbol{\rho}$ for a given $K$. More specifically, we consider balanced codes, i.e., $\boldsymbol{\rho}=\frac{1}{4}[1,1,1,1]^{\mathsf{T}}$ for $L=4$ and $\boldsymbol{\rho}=\frac{1}{8}[1,1,1,1,1,1,1,1]^{\mathsf{T}}$ for $L=8$, as well as two examples of unbalanced codes, i.e., $\boldsymbol{\rho}=\frac{1}{8}[4,2,1,1]^{\mathsf{T}}$ for $L=4$ and $\boldsymbol{\rho}=\frac{1}{16}[4,3,2,2,2,1,1,1]^{\mathsf{T}}$ for $L=8$. From Fig.~\ref{Fig:RateKmulti}, we observe that for fixed $K$ and $L$, balanced codes achieve a higher code rate than unbalanced codes, as expected. Moreover, we observe from Fig.~\ref{Fig:RateKmulti} that increasing the number of symbols decreases the code rate of balanced SCW codes. Nevertheless, as discussed in Section~IV-D, the data rate increases as $L$ increases. Furthermore, for large $K$, the rates approach  the asymptotic bound in Proposition~\ref{Prop:CodeRate}.

In Fig.~\ref{Fig:RateK}, we plot the code rate for binary CW codes, $R^{\mathrm{code}}(K,\omega)$, versus the codeword length, $K$, for different $\rho \in\{\frac{1}{2},\frac{1}{3},\frac{1}{4}\}$. Moreover, we plot the lower and upper bounds presented in Corollary~\ref{Corol:RateBin}. Fig.~\ref{Fig:RateK} reveals that the proposed bounds are quite accurate for all values of $K$ and specifically become very accurate as $K\to\infty$. Moreover,  we observe from Fig.~\ref{Fig:RateK} that the code rate decreases for binary CW codes  as the weight of the code decreases. This is true for any binary CW code if $\rho\leq 0.5$. Furthermore, as $K$ increases, the code rates approach the asymptotic bound given in Corollary~\ref{Corol:RateBin}.

\begin{figure}[!tbp]
  \centering
  \begin{minipage}[b]{0.49\textwidth}
  \centering
\resizebox{1\linewidth}{!}{\psfragfig{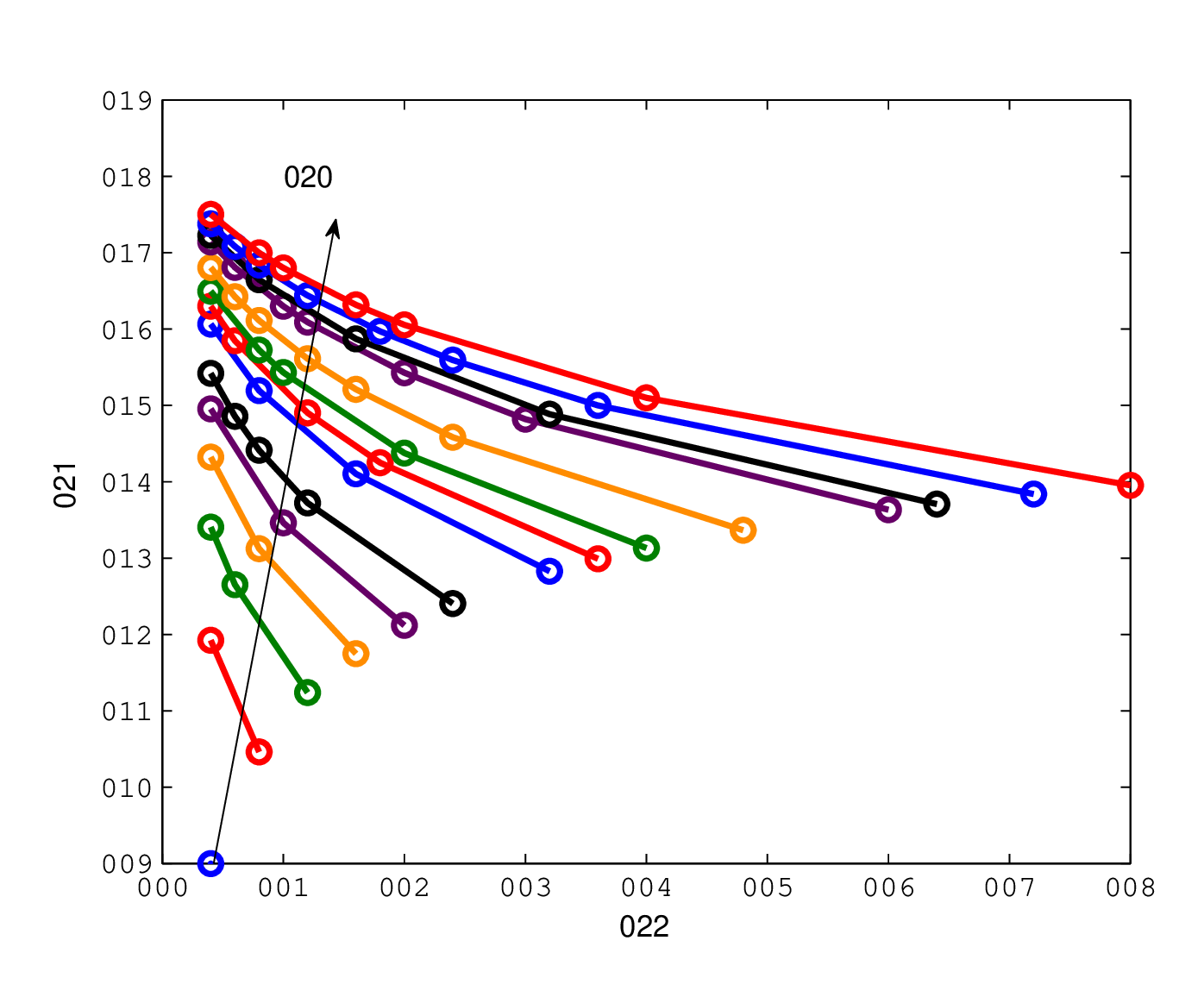}} \vspace{-0.8cm}
\caption{Code rate (unitless) for balanced SCW codes, $R^{\mathrm{code}}(\bar{\boldsymbol{\omega}})$, versus the  number of possible symbols, $L$, for different codeword lengths.\vspace{-0.3cm} }
\label{Fig:RateL}
  \end{minipage}
    \hfill
  \begin{minipage}[b]{0.01\textwidth}
  \end{minipage}
  \hfill
  \begin{minipage}[b]{0.49\textwidth}
  \centering
\resizebox{1\linewidth}{!}{\psfragfig{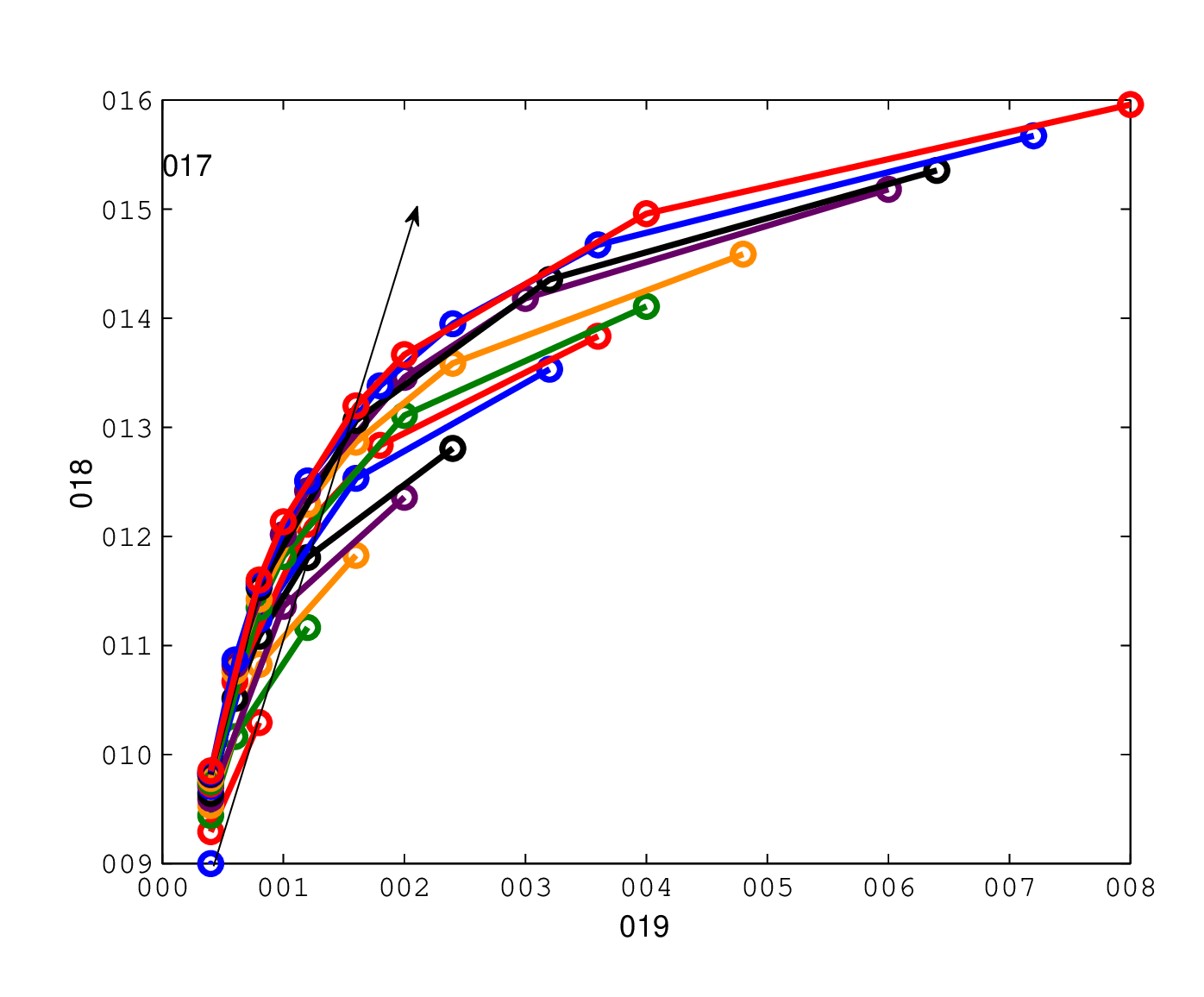}} \vspace{-0.8cm}
\caption{Data rate in bits/symbol for balanced SCW codes, $R^{\mathrm{inf}}(\bar{\boldsymbol{\omega}})$, versus the  number of possible symbols, $L$, for different codeword lengths.\vspace{-0.3cm} }
\label{Fig:RateLinf}
  \end{minipage}
    \hfill
  \begin{minipage}[b]{0.01\textwidth}
  \end{minipage}\vspace{-0.4cm}
\end{figure}

Next, we study the monotonicity of the code rate and the data rate in $L$ as discussed in Section~IV-D. In particular, in Figs.~\ref{Fig:RateL} and \ref{Fig:RateLinf}, we show the code rate and the data rate versus the number of symbols, $L$, for different codeword lengths, $K$, respectively. Note that given $K$, the applicable $L$ has to satisfy the condition that $K/L$ is an integer number. We observe from Fig.~\ref{Fig:RateL} that the code rate is a monotonically decreasing function of $L$. On the contrary, Fig.~\ref{Fig:RateLinf} reveals that the data rate is a monotonically increasing function of $L$. The reason for the different behaviors of the code rate and the data rate is that the code rate specifies the average information content of a codeword compared to uncoded transmission with the same~symbol~set (unitless) whereas the data rate specifies the average information content of the codeword per symbol (in bits per symbol).

\subsubsection{Error Analysis} 

\begin{figure}[!tbp]
  \centering
  \begin{minipage}[b]{0.49\textwidth}
  \centering
\resizebox{1\linewidth}{!}{\psfragfig{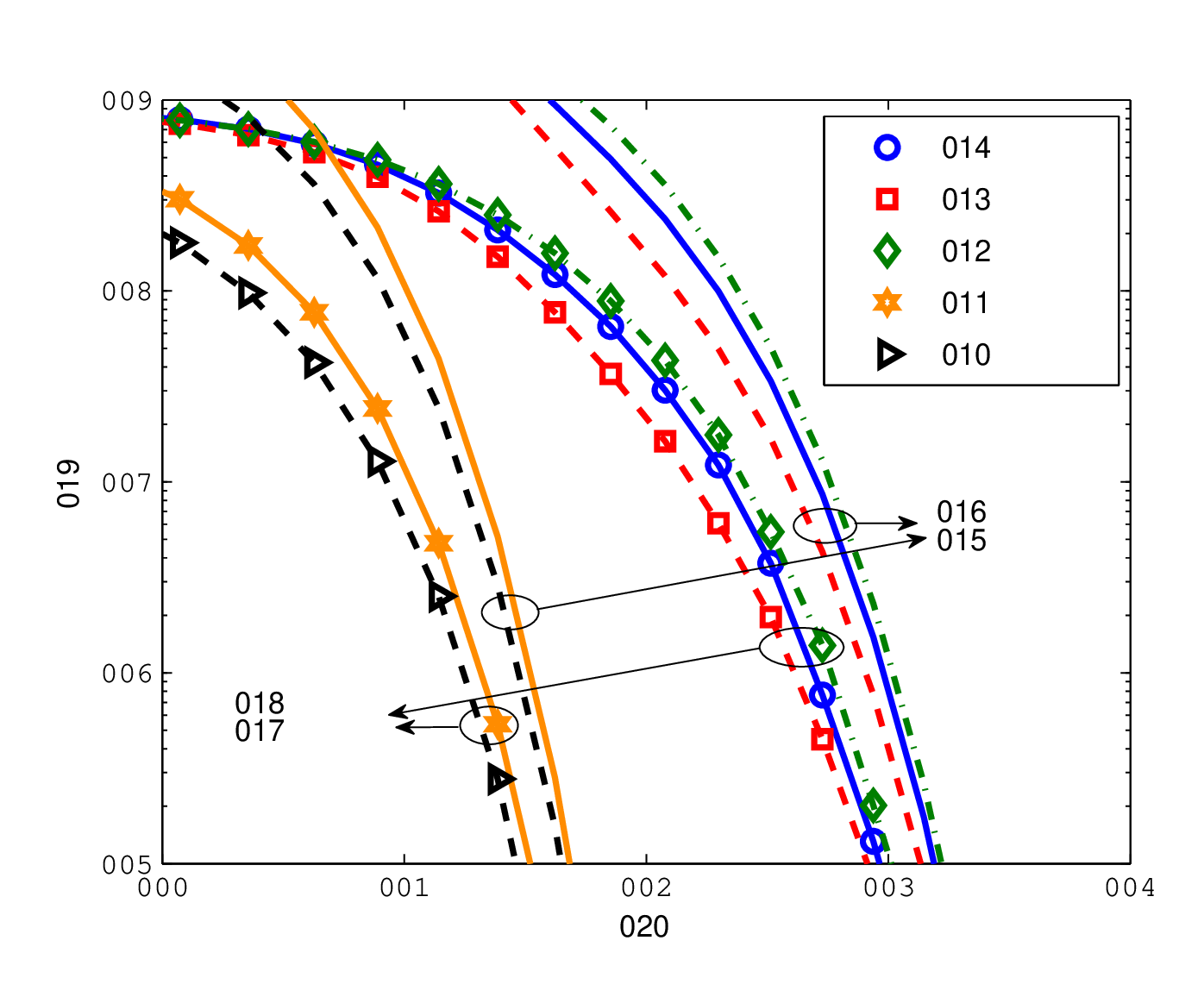}} \vspace{-0.8cm}
\caption{CER for SCW codes, $P_e^{\mathrm{code}}(\bar{\boldsymbol{\omega}}|\bar{\mathbf{c}})$, versus the SINR in dB for different weights $\bar{\boldsymbol{\omega}}$. \vspace{-0.3cm} }
\label{Fig:CER_SCW}
  \end{minipage}
    \hfill
  \begin{minipage}[b]{0.01\textwidth}
  \end{minipage}
  \hfill
  \begin{minipage}[b]{0.49\textwidth}
  \centering
\resizebox{1\linewidth}{!}{\psfragfig{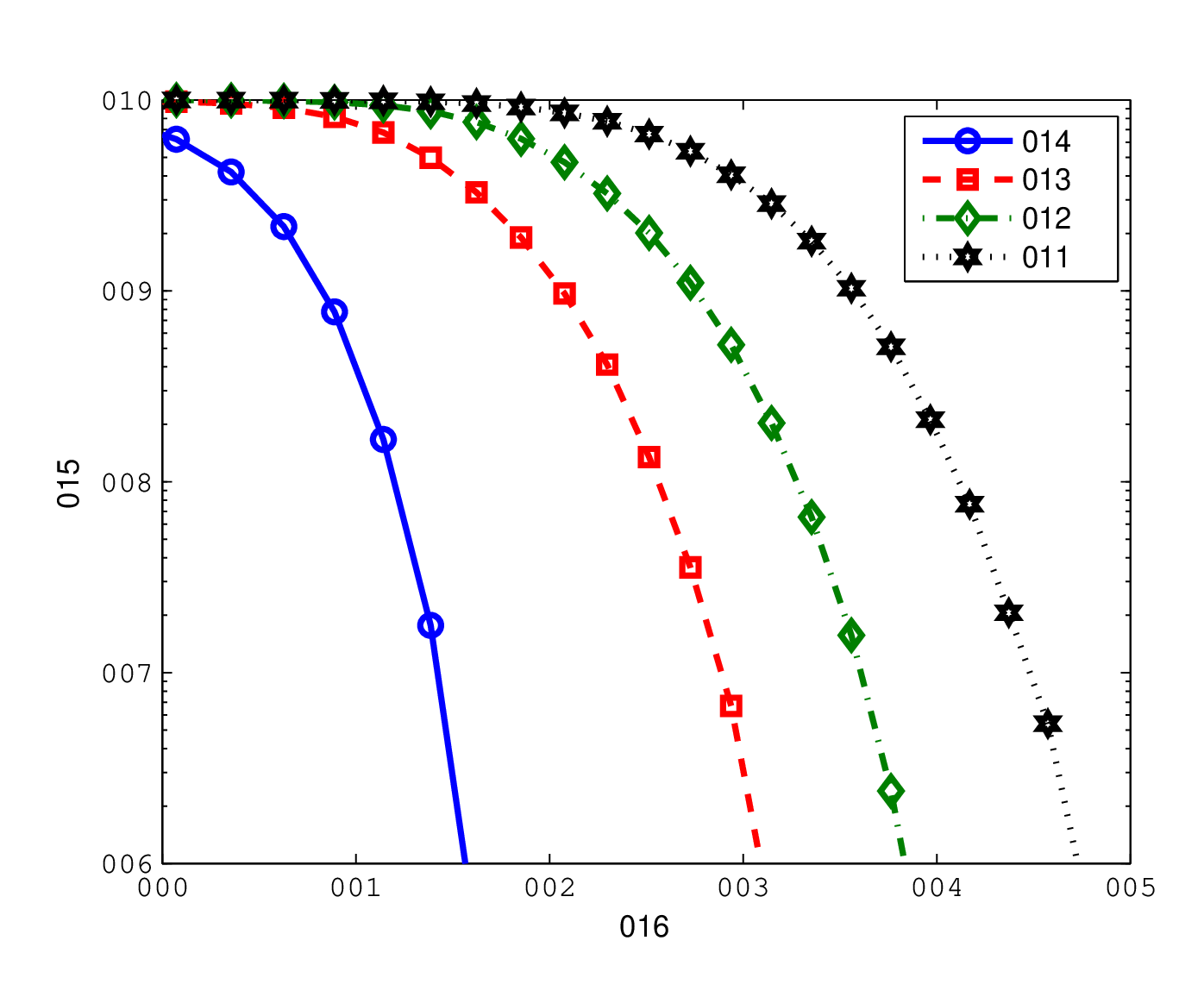}} \vspace{-0.8cm}
\caption{CER for balanced SCW codes, $P_e^{\mathrm{code}}(\bar{\boldsymbol{\omega}}|\bar{\mathbf{c}})$, versus the SINR in dB for $K=12$, and different numbers of symbols $L\in\{2,3,4,6\}$. \vspace{-0.3cm} }
\label{Fig:CER_SCW_Bal}
  \end{minipage}
    \hfill
  \begin{minipage}[b]{0.01\textwidth}
  \end{minipage}\vspace{-0.4cm}
\end{figure}

\begin{figure} 
  \centering
\resizebox{0.5\linewidth}{!}{\psfragfig{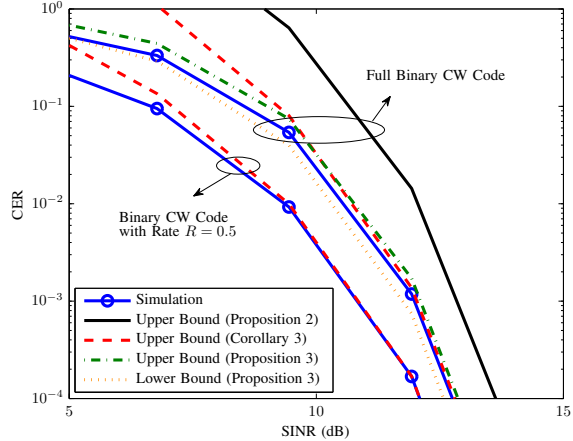}} \vspace{-0.3cm}
\caption{CER for binary CW codes, $P_e^{\mathrm{code}}(K,\omega|\bar{\mathbf{c}})$, versus the SINR in dB for $K=10$ and $\rho=\frac{1}{2}$.\vspace{-0.3cm} }
\label{Fig:CER_SNR}
\end{figure}

In the following, we evaluate the error performance of the proposed CSI-free detector. To examine the performance of different SCW codes, we adopt a simple ternary symbol set, i.e., $\mathcal{S}=\{0,0.5,1\}$, and a codeword length of $K=6$. Moreover, we consider the following five weight vectors: $\bar{\boldsymbol{\omega}}=[2,2,2]^{\mathsf{T}}$ which yields a balanced code, $\bar{\boldsymbol{\omega}}=[3,2,1]^{\mathsf{T}},[1,2,3]^{\mathsf{T}}$ which yield unbalanced codes, $\bar{\boldsymbol{\omega}}=[3,0,3]^{\mathsf{T}}$ which is equivalent to a binary balanced code, and $\bar{\boldsymbol{\omega}}=[5,0,1]^{\mathsf{T}}$ which is equivalent to pulse position modulation (PPM) \cite{PPM_Pieroborn}. In Fig.~\ref{Fig:CER_SCW}, we show the CER for these SCW codes, $P_e^{\mathrm{code}}(\bar{\boldsymbol{\omega}}|\bar{\mathbf{c}})$, versus the SINR in dB. In addition, we plot the upper bound given in Proposition~\ref{Prop:UpperGen} for $t=0.5$ \footnote{For simplicity, we choose a fixed $t$ for the results shown in Figs.~\ref{Fig:CER_SCW} and \ref{Fig:CER_SNR}, i.e., $t=0.5$. Moreover, this specific value of $t$ was chosen in a trial-and-error manner  without claim of optimality of the chosen~$t$.}. Fig.~\ref{Fig:CER_SCW} confirms the validity of the proposed upper bound and that it becomes tighter at high SINRs. We note that all codes considered in Fig.~\ref{Fig:CER_SCW} do not require CSI for detection, have identical codeword length, $K$, have the same identical  per-symbol ``power" constraint, $N^{\mathrm{tx}}$, and \textit{in principle} employ the same symbol set, $\mathcal{S}$. However, their code rates, $R^{\mathrm{code}}(\bar{\boldsymbol{\omega}})$, and average power consumptions, $\bar{N}^{\mathrm{tx}}(\bar{\boldsymbol{\omega}})$, are not necessarily identical, which makes a direct performance comparison difficult. 
Therefore, in Fig.~\ref{Fig:CER_SCW_Bal}, we show the CER versus the SINR only  for balanced SCW codes with $K=12$ and different numbers of symbols $L\in\{2,3,4,6\}$.  Since all balanced SCW codes have identical average energy consumption, i.e.,  $\bar{N}^{\mathrm{tx}}(\bar{\boldsymbol{\omega}})=\frac{1}{2}N^{\mathrm{tx}}$, cf. Section~V-D, the only difference between the curves in Fig.~\ref{Fig:CER_SCW_Bal} is their achievable code rate/data rate. From Fig.~\ref{Fig:CER_SCW_Bal}, we observe that as $L$ decreases, the CER performance improves at the expense of a lower data rate.

The SCW codes adopted for Fig.~\ref{Fig:CER_SCW} are full codes, i.e., all possible codewords are used. In Corollary~\ref{Corol:BinaryCW}, we showed that CSI-free detection is possible also for binary CW codes with partial codebooks. In Fig.~\ref{Fig:CER_SNR}, we show the CER for binary CW codes, $P_e^{\mathrm{code}}(K,\omega|\bar{\mathbf{c}})$, versus the SINR in dB for $K=10$ and $\rho=\frac{1}{2}$. Results for both the partial code with code rate $R=0.5$ and the full code with rate $R(K,\omega)=0.8$ are included. In particular, to generate the partial codebook, $2^{0.5K}=32$ codewords are randomly chosen out of all $M=252$ possible codewords. We observe that the code with partial codebook achieves a lower CER at the expense of a lower code rate. In addition, in Fig.~\ref{Fig:CER_SNR}, we show the upper bounds proposed in Proposition~\ref{Prop:UpperGen},  Corollary~\ref{Corol:CER_UppBin}, and Proposition~\ref{Prop:UppBinFull} and the  lower bound  proposed in Proposition~\ref{Prop:UppBinFull}. We note that  the bounds in Proposition~\ref{Prop:UppBinFull} are valid only for full codes. Fig.~\ref{Fig:CER_SNR} confirms the validity of the bounds and that the upper bounds proposed in  Proposition~\ref{Prop:UppBinFull} and  Corollary~\ref{Corol:CER_UppBin} for the binary CW codes are tighter than the upper bound proposed in Proposition~\ref{Prop:UpperGen} for general SCW codes. Moreover, Fig.~\ref{Fig:CER_SNR} reveals that the bounds in Proposition~\ref{Prop:UppBinFull} are fairly tight for all SINRs whereas the upper bound in Corollary~\ref{Corol:CER_UppBin} is particularly tight at high SINRs.

\subsubsection{Trade-Off and Performance Comparison} 
\begin{figure}[!tbp]
  \centering
  \begin{minipage}[b]{0.32\textwidth}
  \centering
\resizebox{1.1\linewidth}{!}{\psfragfig{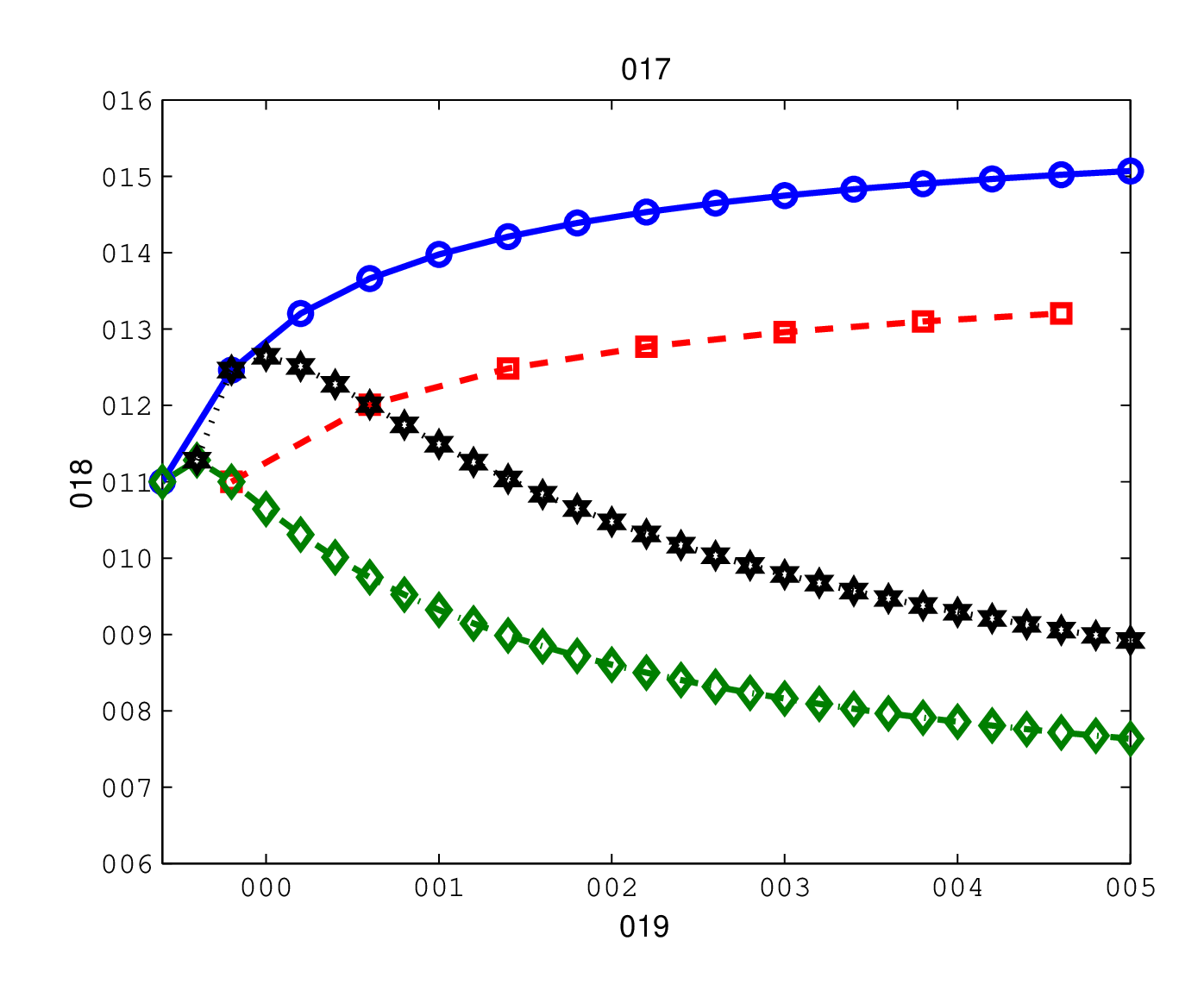}} \vspace{-0.3cm}
  \end{minipage}
    \hfill
  \begin{minipage}[b]{0.01\textwidth}
  \end{minipage}
  \hfill
  \begin{minipage}[b]{0.32\textwidth}
  \centering
\resizebox{1.1\linewidth}{!}{\psfragfig{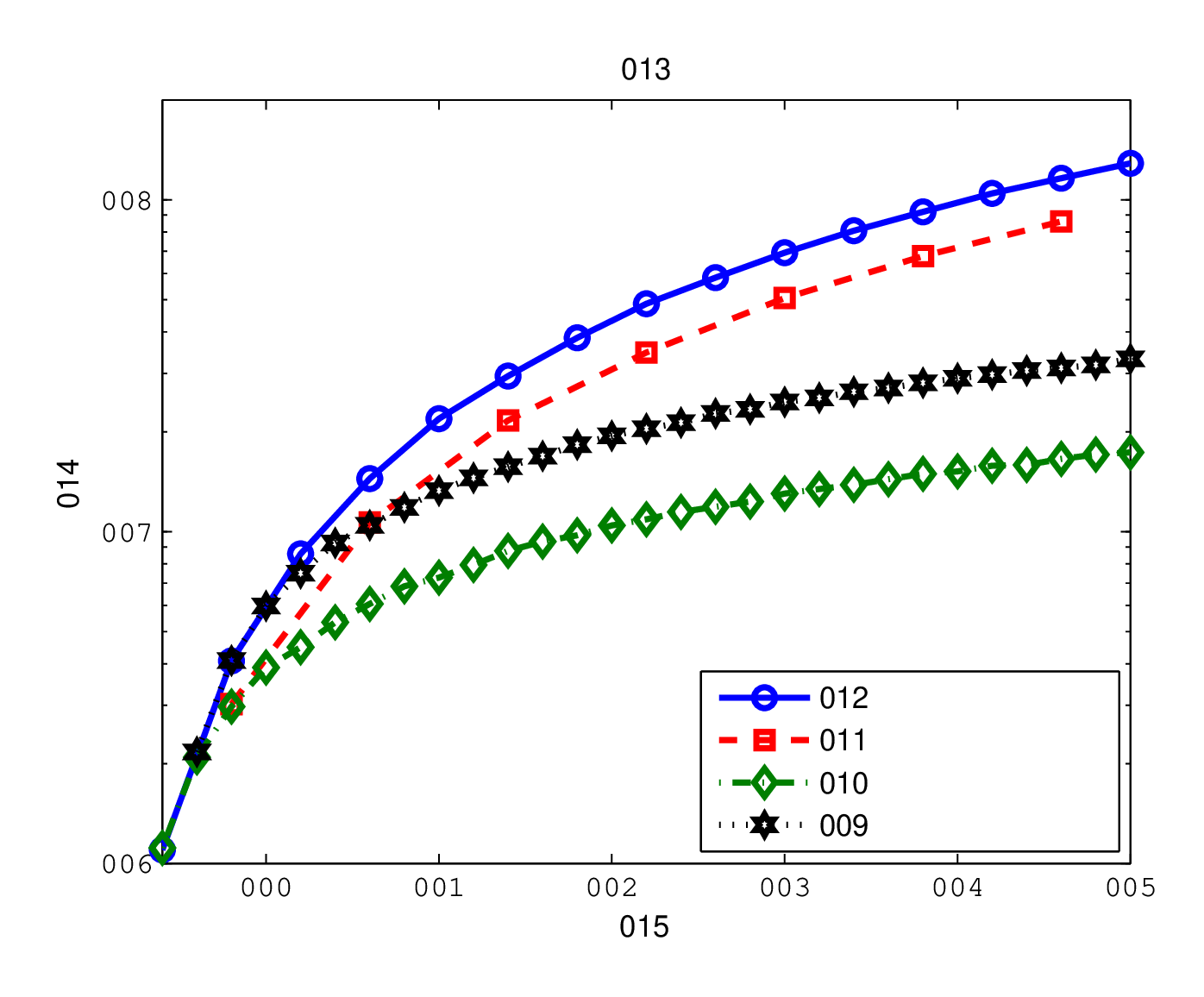}} \vspace{-0.3cm}
  \end{minipage}
    \hfill
      \begin{minipage}[b]{0.32\textwidth}
  \centering
\resizebox{1.1\linewidth}{!}{\psfragfig{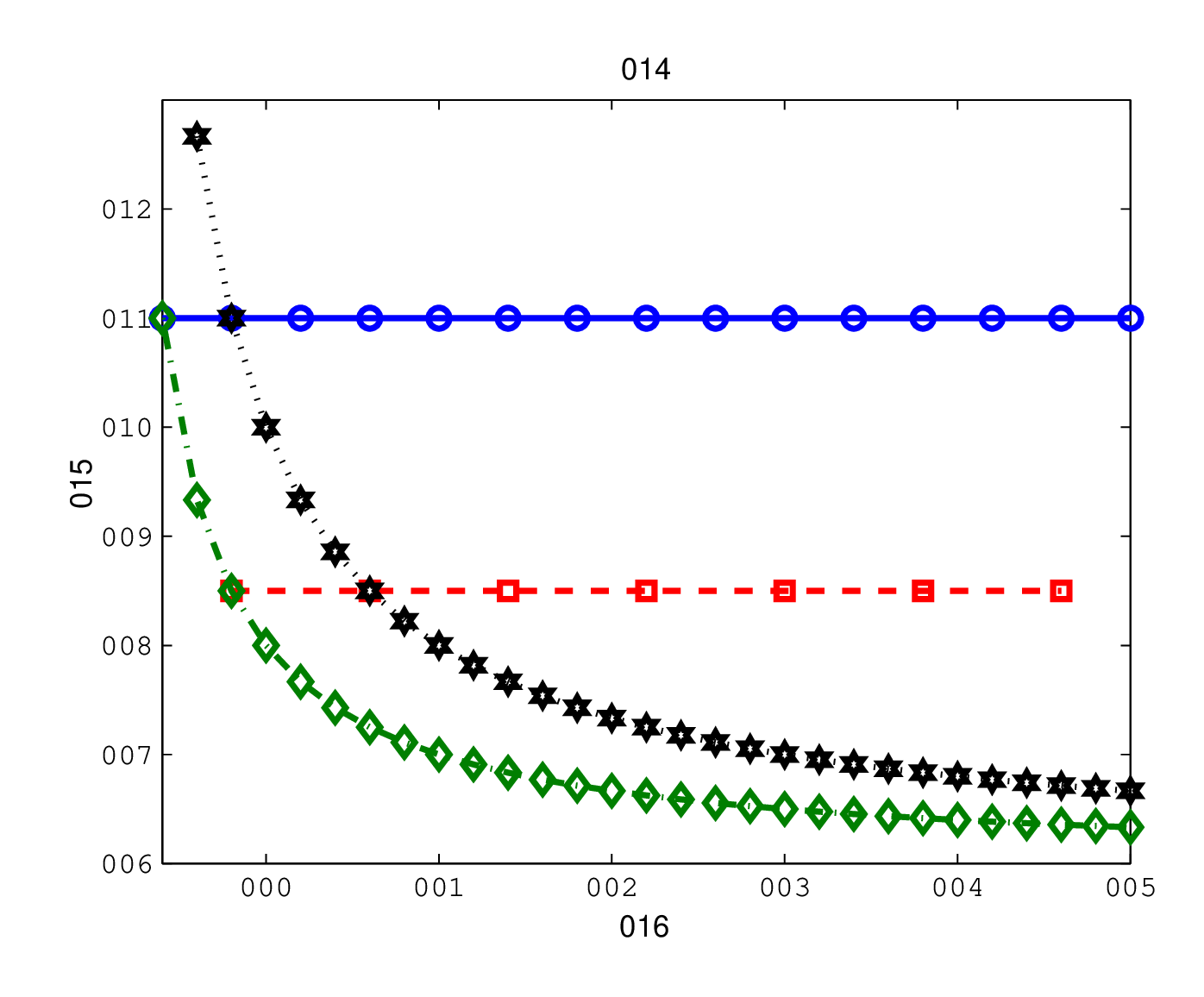}} \vspace{-0.3cm}
  \end{minipage}
    \hfill
  \begin{minipage}[b]{0.01\textwidth}
  \end{minipage}\vspace{-0.4cm}
  \caption{a) Data rate, $R^{\mathrm{inf}}(K,\omega)$,  b) CER, $P_e^{\mathrm{code}}(K,\omega|\bar{\mathbf{c}})$, and c) normalized average number of released molecules, $\bar{N}^{\mathrm{tx}}_{\mathrm{nrm}}(K,\omega)$, for binary CW codes versus the codeword length, $K$,  for $\mathsf{SINR}=10$ dB. \vspace{-0.3cm} }
\label{Fig:Tradeoff}
\vspace{-0.4cm}
\end{figure}

In order to reveal the full trade-off between  rate, error performance, and the average number of released molecules, in Figs.~\ref{Fig:Tradeoff} a), b), and c), we show respectively the data rate, $R^{\mathrm{inf}}(K,\omega)$, the CER, $P_e^{\mathrm{code}}(K,\omega|\bar{\mathbf{c}})$, and the normalized average number of released molecule, $\bar{N}^{\mathrm{tx}}_{\mathrm{nrm}}(K,\omega)=\frac{\bar{N}^{\mathrm{tx}}(\bar{\boldsymbol{\omega}})}{N^{\mathrm{tx}}}$  of binary CW codes versus the codeword length, $K$,  for $\mathsf{SINR}=10$ dB. We consider  four binary CW codes:, namely the balanced code with weight $\omega=\frac{K}{2}$, an unbalanced code with weight $\omega=\frac{K}{4}$, PPM, i.e., the CW code with weight $\omega=1$, and MPPM with two pulses, i.e., the CW code with weight $\omega=2$. From Fig.~\ref{Fig:Tradeoff} a), we observe that the data rates of the considered balanced and unbalanced CW codes  increase with increasing $K$, whereas the data rates of PPM and MPPM decrease for large $K$.  In Fig.~\ref{Fig:Tradeoff} c), the normalized average energy consumptions of the considered balanced and unbalanced CW codes  are constant for all $K$, whereas the normalized average energy consumptions of PPM and MPPM decrease with increasing $K$. In Fig.~\ref{Fig:Tradeoff} b), we observe that the CERs of all the considered CW codes increases with increasing $K$. In total, from Fig.~\ref{Fig:Tradeoff}, we observe that the following relations hold for large~$K$ 
\begin{IEEEeqnarray}{ccccccc} 
R^{\mathrm{inf}}(K,K/2)&>&R^{\mathrm{inf}}(K,K/4)&>&R^{\mathrm{inf}}(K,2)&>&R^{\mathrm{inf}}(K,1) \nonumber \\
P_e^{\mathrm{code}}(K,K/2|\bar{\mathbf{c}})&>&P_e^{\mathrm{code}}(K,K/4|\bar{\mathbf{c}})&>&P_e^{\mathrm{code}}(K,2|\bar{\mathbf{c}})&>&P_e^{\mathrm{code}}(K,1|\bar{\mathbf{c}}) \nonumber\\
\bar{N}^{\mathrm{tx}}_{\mathrm{nrm}}(K,K/2)&>&\bar{N}^{\mathrm{tx}}_{\mathrm{nrm}}(K,K/4)&>&\bar{N}^{\mathrm{tx}}_{\mathrm{nrm}}(K,2)&>&\bar{N}^{\mathrm{tx}}_{\mathrm{nrm}}(K,1). \nonumber
\end{IEEEeqnarray}

\begin{figure} 
  \centering
\resizebox{0.5\linewidth}{!}{\psfragfig{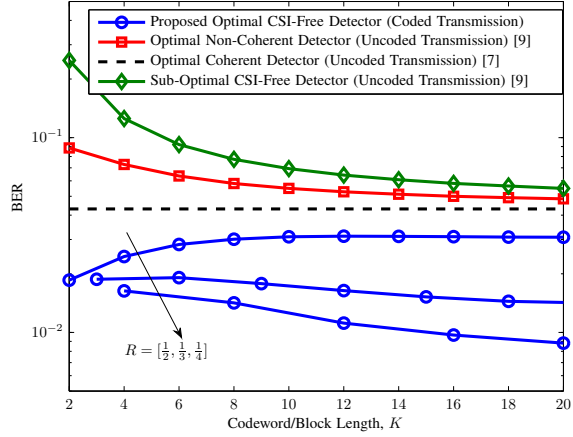}} \vspace{-0.3cm}
\caption{BER versus the codeword/block length, $K$, for $\rho=\frac{1}{2}$, $\mathrm{SINR}=10$ dB, and $R\in\{\frac{1}{2},\frac{1}{3},\frac{1}{4}\}$. The code rate decreases along the direction of the arrow for the proposed CSI-free detectors. \vspace{-0.3cm} }
\label{Fig:BER_SNR}
\end{figure}

In Fig.~\ref{Fig:BER_SNR}, we consider the stochastic channel model introduced in \cite{NanoCOM16} and compare the proposed coded communication scheme with uncoded transmission employing the coherent symbol-by-symbol detector in \cite{HamidJSAC} and the optimal non-coherent and the sub-optimal CSI-free detectors in \cite{NanoCOM16}.  In Fig.~\ref{Fig:BER_SNR},  we show the BER versus the codeword/block length, $K$, for $\rho=\frac{1}{2}$, $\mathsf{SINR}=10$ dB, $R\in\{\frac{1}{2},\frac{1}{3},\frac{1}{4}\}$ and Scenario~2 of the stochastic MC channel in \cite{NanoCOM16}. The BERs of the optimal non-coherent and the sub-optimal CSI-free detectors approach that of the optimal coherent detector as $K\to\infty$. The proposed CSI-free detector based on SCW codes outperforms all considered uncoded benchmark schemes at the expense of a lower data rate. Furthermore, the gain of the proposed coded communication over the uncoded benchmark schemes increases as the code rate decreases. The BER curves for the proposed SCW codes are not necessarily monotonic in $K$. In fact, for a full code, as $K$ increases, we expect the CER to increase\footnote{For full SCW codes and assuming codeword $\mathbf{s}$ is transmitted and observation vector $\mathbf{r}$ is received, an error occurs if there exist $k$ and $k'$ for which $r[k]>r[k']$ and $s[k]<s[k']$ hold. Therefore, the probability of this error event increases for larger $K$ which leads to the monotonically increasing behavior of CER with respect to $K$.}.  However, this may not be valid for a code with a given rate where only a subset of all available codewords is adopted. In addition, the relation between CER and BER is influenced by the adopted bit-sequence-to-codeword mapping. Therefore, the BER depends on the codebook selection and the bit-sequence-to-codeword mapping strategy. For the rates considered in Fig.~\ref{Fig:BER_SNR}, we observe that as $K$ increases, the BER increases for $R=\frac{1}{2}$ and decreases for $R=\frac{1}{3},\frac{1}{4}$.

\section{Conclusions and Future Work}

In this paper, we proposed SCW codes which facilitate \textit{optimal ML CSI-free} sequence detection at the expense of a decrease in the data rate compared to uncoded transmission.  We analyzed the code rate, the error rate, and the average number of released molecules for general SCW codes. In addition, we studied the properties of binary SCW codes and balanced~SCW~codes in further detail. Simulation results verified our analytical derivations and showed that the proposed SCW codes with CSI-free detection  outperform uncoded transmission with optimal coherent and non-coherent detection. 
 
The framework developed in this paper can be extended in several directions. First, for symbols with binary alphabet, we showed that CSI-free ML sequence detection is possible for both full and partial codebooks, cf. Corollary~\ref{Corol:BinaryCW}. It is of interest to develop a systematic approach for selecting the codewords for the partial codebook such that the average or minimum  distance between the codewords is maximized and thereby the BER is minimized \cite{CodeMinDis}. Second, for the simulation results provided in Section~V, a random bit-sequence-to-codeword mapping was employed for simplicity. The systematic design of mapping strategies, which ensure that the codeword pairs with the highest pairwise error probability are mapped to bit sequences that have the minimum possible Hamming distance, and thereby minimize the BER, is an interesting research problem.

\appendices

\section{Proof of Lemma~\ref{Lem:ML}}\label{App:LemML}

The ML problem in (\ref{Eq:ML}) can be rewritten as
\begin{align} 
  \hat{\mathbf{s}}  & = \underset{\mathbf{s}\in\boldsymbol{\mathcal{S}}}{\mathrm{argmax}} \,\, 
  \prod_{{\ell}=0}^{L-1} \left( \eta_{\ell} \bar{c}_{\mathrm{s}} + \bar{c}_{\mathrm{n}} \right)^{\sum_{k=1}^K r[k] \mathbf{1}\{s[k]=\eta_{\ell}\}} 
   \frac{ \mathsf{exp}\left(- \bar{c}_{\mathrm{s}} \sum_{k=1}^K  s[k] - K \bar{c}_{\mathrm{n}} \right)}{\prod_{k=1}^{K} r[k]!}, \nonumber \\
   & \overset{(a)}{=} \underset{\mathbf{s}\in\boldsymbol{\mathcal{S}}}{\mathrm{argmax}} \,\, 
   \prod_{\ell=0}^{L-1} \left( 1+ \eta_{\ell} \mathsf{SIR}  \right)^{\omega_{\ell}(\mathbf{s},\mathbf{r})} \mathsf{exp}\left(- \omega(\mathbf{s}) \bar{c}_{\mathrm{s}} \right)  
   \frac{\prod_{\ell=0}^{L-1}{\bar{c}_{\mathrm{n}}}^{\omega_{\ell}(\mathbf{s},\mathbf{r})} \mathsf{exp}\left(- K \bar{c}_{\mathrm{n}} \right)}{\prod_{k=1}^{K} r[k]!} \nonumber \\
   & \overset{(b)}{=} \underset{\mathbf{s}\in\boldsymbol{\mathcal{S}}}{\mathrm{argmax}} \,\, 
   \prod_{\ell=0}^{L-1} \left( 1+ \eta_{\ell} \mathsf{SIR}  \right)^{\omega_{\ell}(\mathbf{s},\mathbf{r})} \mathsf{exp}\left(- \omega(\mathbf{s}) \bar{c}_{\mathrm{s}} \right)  
   \frac{{\bar{c}_{\mathrm{n}}}^{\sum_{k=1}^{K}r[k]} \mathsf{exp}\left(- K \bar{c}_{\mathrm{n}} \right)}{\prod_{k=1}^{K} r[k]!},
\end{align}
where for equality $(a)$, we use definitions $\omega_{\ell}(\mathbf{s},\mathbf{r})=\sum_{k=1}^K r[k] \mathbf{1}\{s[k]=\eta_{\ell}\}$ and $\omega(\mathbf{s})=\sum_{k=1}^K  s[k]$, and for equality $(b)$, we use the identity $\prod_{\ell=0}^{L-1}{\bar{c}_{\mathrm{n}}}^{\omega_{\ell}(\mathbf{s},\mathbf{r})}={\bar{c}_{\mathrm{n}}}^{\sum_{k=1}^{K}r[k]}$. Note that the term $\frac{{\bar{c}_{\mathrm{n}}}^{\sum_{k=1}^{K}r[k]} \mathsf{exp}\left(- K \bar{c}_{\mathrm{n}} \right)}{\prod_{k=1}^{K} r[k]!}$ does not depend on the hypothesis sequence and hence, cannot change the ML solution. Therefore, the ML problem can be simplified as
\begin{align} 
  \hat{\mathbf{s}}  & = \underset{\mathbf{s}\in\boldsymbol{\mathcal{S}}}{\mathrm{argmax}} \,\, 
   \mathsf{exp}\left(- \omega(\mathbf{s}) \bar{c}_{\mathrm{s}} \right)  \prod_{\ell=0}^{L-1} \left( 1+ \eta_{\ell} \mathsf{SIR}  \right)^{\omega_{\ell}(\mathbf{s},\mathbf{r})} \nonumber \\
   & \overset{(a)}{=} \underset{\mathbf{s}\in\boldsymbol{\mathcal{S}}}{\mathrm{argmax}} \,\, 
   - \omega(\mathbf{s}) \bar{c}_{\mathrm{s}}  + \sum_{\ell=1}^{L-1} \omega_{\ell}(\mathbf{s},\mathbf{r}) \mathsf{ln}\left( 1+ \eta_{\ell} \mathsf{SIR}  \right) 
   \triangleq \Lambda^{\mathrm{ML}}(\mathbf{s}),
\end{align}
where to arrive at equality $(a)$, we use the property that $\mathsf{ln}(\cdot)$ is a monotonically increasing function and we removed index $\ell=0$ in the summation as it leads to $\mathsf{ln}\left( 1+ \eta_{\ell} \mathsf{SIR}  \right)=0$ for $\eta_{0}=0$. The above solution is given in Lemma~\ref{Lem:ML} which concludes the proof.

\section{Proof of Theorem~\ref{Theo:ML_StrCW}}
  \label{App:ML_StrCW}

For SCW codes, $\omega(\mathbf{s})$ is identical for all codewords and hence does not change the ML sequence. Therefore, the coherent ML problem in (\ref{Eq:ML_Sol}) simplifies to
\begin{align}  \label{Eq:ML_StrCW_Eq}
  \hat{\mathbf{s}}  & = \underset{\mathbf{s}\in\boldsymbol{\mathcal{S}}^{\mathrm{sc}}(\bar{\boldsymbol{\omega}})}{\mathrm{argmax}} \,\,  
    \sum_{{\ell}=1}^{L-1} \omega_{\ell}(\mathbf{s},\mathbf{r}) \mathsf{ln}\left( 1+ \eta_{\ell} \mathsf{SIR}  \right) \nonumber \\
      & = \underset{\mathbf{s}\in\boldsymbol{\mathcal{S}}^{\mathrm{sc}}(\bar{\boldsymbol{\omega}})}{\mathrm{argmax}} \,\,  
    \sum_{k=1}^{K} r[k] \mathsf{ln}\left( 1+ s[k] \mathsf{SIR}  \right).
\end{align}
 The expression in (\ref{Eq:ML_StrCW_Eq}) is in fact  
a weighted sum of the observations $r[k]$ where the weights $\mathsf{ln}\left( 1+ s[k] \mathsf{SIR}  \right)$ are monotonically increasing functions of $s[k]$. Therefore, for the ML sequence $\mathbf{s}^*=[s^*[1],\dots,s^*[k]]^{\mathsf{T}}$, if $r[k]\geq r[k']$  holds, then $s^*[k]\geq s^*[k']$ has to hold. This leads to Algorithm~1 for general SCW codes.  For the case of binary CW codes, $\boldsymbol{\mathcal{S}}^{\mathrm{c}}(K,\omega)$, this leads to a sequence whose ``1" elements  correspond to the $\omega$ largest elements of $\mathbf{r}$.  The resulting sequence is optimal if it belongs to the codebook $\boldsymbol{\mathcal{S}}^{\mathrm{sc}}(\bar{\boldsymbol{\omega}})$. This condition is ensured if the code is full. 
Note that this is the solution of the ML problem in (\ref{Eq:ML}) for coherent sequence detection. If for a given CSI $(\bar{c}_{\mathrm{s}},\bar{c}_{\mathrm{n}})$, the sequence $\mathbf{s}^*$  that maximizes the conditional PDF $f_{\mathbf{r}}(\mathbf{r}|\bar{\mathbf{c}},\mathbf{s})$ does not depend on the CSI value, the average PDF $\mathsf{E}_{\bar{\mathbf{c}}}\{f_{\mathbf{r}}(\mathbf{r}|\bar{\mathbf{c}},\mathbf{s})\}$ in (\ref{Eq:ML_NonCoherent}) is also maximized by $\mathbf{s}^*$. In other words, the solutions of (\ref{Eq:ML}) and (\ref{Eq:ML_NonCoherent}) for coherent and non-coherent detection are identical and do not depend on instantaneous nor statistical CSI. Therefore, an SCW code enables optimal CSI-free detection.  These results are concisely summarized in Theorem~\ref{Theo:ML_StrCW} and Algorithm~1 which concludes~the~proof.

\section{Proof of Proposition~\ref{Prop:CodeRate}}\label{App:PropRate}

In the following, using (\ref{Eq:CodeRateGen}), we derive the code rate of a \textit{full} SCW code. First, note that we have $|\mathcal{S}|=L$ and $K=\sum_{\ell=0}^{L-1}\bar{\omega}_{\ell}$ for SCW code $\boldsymbol{\mathcal{S}}^{\mathrm{sc}}(\bar{\boldsymbol{\omega}})$. In order to determine the number of codewords $M$ for a given SCW code $\boldsymbol{\mathcal{S}}^{\mathrm{sc}}(\bar{\boldsymbol{\omega}})$, we use the definition of the binomial coefficient, i.e., ${n\choose k}=\frac{n!}{k!(n-k)!}$. In particular, there are ${K\choose \bar{\omega}_{L-1}}$ possibilities for the positions of symbol $\eta_{L-1}=1$. Having fixed the positions of symbol $\eta_{L-1}$, there are ${K-\bar{\omega}_{L-1}\choose \bar{\omega}_{L-2}}$ possibilities for the positions of symbol $\eta_{L-2}$. Continuing this process, we obtain $M$ for a full SCW code $\boldsymbol{\mathcal{S}}^{\mathrm{sc}}(\bar{\boldsymbol{\omega}})$ as
\begin{IEEEeqnarray}{lll} \label{Eq:StrCT_M}
M&={K\choose \bar{\omega}_{L-1}}{K-\bar{\omega}_{L-1}\choose \bar{\omega}_{L-2}}
\cdots{\bar{\omega}_{0}+\bar{\omega}_{1}\choose \bar{\omega}_{1}} {\bar{\omega}_{0}\choose \bar{\omega}_{0}} \nonumber \\
&=\prod_{\ell =0}^{L-1} {\sum_{\ell'\leq\ell} \bar{\omega}_{\ell'}\choose \bar{\omega}_{\ell}} 
= \frac{K!}{\prod_{\ell=0}^{L-1}\bar{\omega}_{\ell}!}.
\end{IEEEeqnarray}
Substituting the above results into (\ref{Eq:CodeRateGen}) leads to the first expression in (\ref{Eq:StrCT_Rate}). We note that the first expression in (\ref{Eq:StrCT_M}) is the well-known multinomial coefficient which can be written equivalently as the second expression in (\ref{Eq:StrCT_M}) \cite{TableIntegSerie}. Finally, we note that the entropy of an RV with multinomial distribution and probability vector $\boldsymbol{\rho}=[\rho_0,\rho_1,\dots,\rho_L]^{\mathsf{T}}$ where $\rho_{\ell}=\bar{\omega}_{\ell}/K$, asymptotically approaches $H_L(\boldsymbol{\rho})$ when $K\to\infty$ \cite{TableIntegSerie}. Therefore, we obtain $\mathsf{log}_L(M)\to KH_L(\boldsymbol{\rho})$ as $K\to\infty$. This leads to the asymptotic result in  (\ref{Eq:StrCT_Rate})  and concludes the proof.

\section{Proof of Corollary~\ref{Corol:RateBin}}\label{App:CorolRateBin}

For  full binary CW code $\boldsymbol{\mathcal{S}}^{\mathrm{c}}(K,\omega)$, the number of possible codewords is given by $M={K \choose \omega}$. Therefore, the code rate can be obtained as $R(K,\omega)=\mathsf{log}_2\big({K \choose \omega}\big) /K$. Next, we employ the Stirling approximation of the factorial function given by  \cite{TableIntegSerie}
\begin{IEEEeqnarray}{lll} \label{Eq:Stirling}
 n! = \beta n^{n+0.5} e^{-n},\quad \beta\in[\sqrt{2\pi},e].
\end{IEEEeqnarray}
In particular, substituting the  Stirling approximation into the binomial coefficient, we obtain
\begin{IEEEeqnarray}{lll} \label{Eq:Binomial}
\mathsf{log}_2(M)&=\mathsf{log}_2\left(\frac{K!}{(\rho K)!((1-\rho)K)!}\right) \nonumber \\
&= \mathsf{log}_2\left(\frac{ \beta_1 K^{K+0.5}e^{-K}}{\beta_2 (\rho K)^{\rho K +0.5} e^{-\rho K} \beta_3 ((1-\rho)K)^{(1-\rho)K+0.5} e^{-(1-\rho)K} }\right) \nonumber \\
&= \mathsf{log}_2\left(\frac{\beta_1}{\beta_2\beta_3} \frac{1}{\rho^{\rho K + 0.5}(1-\rho)^{(1-\rho)K+0.5}} \right) \nonumber \\
&= -K [\rho\mathsf{log}_2(\rho)+(1-\rho)\mathsf{log}_2(1-\rho)] - \mathsf{log}_2\left(\frac{\beta_2\beta_3}{\beta_1}\sqrt{\rho(1-\rho)K}\right), \quad
\end{IEEEeqnarray}
where by substituting $\alpha=\frac{\beta_1}{\beta_2\beta_3}$ into the above equation,  we arrive at the second equation in (\ref{Eq:CodeRateBin}) for code rate  $R(K,\omega)=\mathsf{log}_2(M)/K$. Note that since $\beta_1,\beta_2,\beta_3\in[\sqrt{2\pi},e]$ holds, we obtain $\alpha\in[\sqrt{2\pi}/e^2,e/2\pi]$.   This completes the proof.

\section{Proof of Proposition~\ref{Prop:UpperGen}}\label{App:PropUpperGen}

The PEP, denoted by $P(\mathbf{s}\to\hat{\mathbf{s}})$, is defined as the probability that assuming $\mathbf{s}$ is transmitted, $\hat{\mathbf{s}}$ is detected. Using the PEP, the CER is upper bounded based on the union bound as follows
\begin{IEEEeqnarray}{lll} \label{Eq:UB}
P_e^{\mathrm{code}}(\bar{\boldsymbol{\omega}}|\bar{\mathbf{c}}) & \leq \sum_{\forall\mathbf{s}} \sum_{\forall\hat{\mathbf{s}}\neq \mathbf{s}} P(\mathbf{s}\to\hat{\mathbf{s}}) \mathsf{Pr}(\mathbf{s}) \nonumber \\
 &\overset{(a)}{\leq}  \frac{1}{M} \sum_{\forall\mathbf{s}} \sum_{\forall\hat{\mathbf{s}}\neq \mathbf{s}} \mathsf{Pr}\{ X\geq 0\} \nonumber \\
 & \overset{(b)}{\leq} \frac{1}{M} \sum_{\forall\mathbf{s}} \sum_{\forall\hat{\mathbf{s}}\neq \mathbf{s}} 
  \mathsf{E}\left\{\mathsf{exp}\left(Xt\right)\right\} \nonumber \\
 & =\frac{1}{M} \sum_{\forall\mathbf{s}} \sum_{\forall\hat{\mathbf{s}}\neq \mathbf{s}}  G_X(t),\,\,\forall t>0,
\end{IEEEeqnarray}
where in inequality $(a)$, we use the property that the codewords are equiprobable, i.e., $\mathsf{Pr}(\mathbf{s})=\frac{1}{M}$, define $X=\Lambda^{\mathrm{ML}}(\hat{\mathbf{s}}) - \Lambda^{\mathrm{ML}}(\mathbf{s})$, and treat the case $X=0$ always as an error which upper bounds the PEP term $P(\mathbf{s}\to\hat{\mathbf{s}})$. For inequality $(b)$, we employ the Chernoff bound where $G_X(t)$ denotes the moment generating function (MGF) of RV $X$ \cite{StochGallager}. 
\begin{remk}
 Suppose that the adopted SCW code is full. Thereby, due to the symmetry of the codewords, the error probabilities for all codewords are identical and the bound in (\ref{Eq:UB}) can be computed  only for one arbitrarily chosen codeword $\mathbf{s}$. Hence,  the summation over $\mathbf{s}$ is not needed and  the upper bound simplifies to
 $P_e^{\mathrm{code}}(\bar{\boldsymbol{\omega}}|\bar{\mathbf{c}}) \leq \sum_{\forall\hat{\mathbf{s}}\neq \mathbf{s}}  G_X(t),\,\,\forall t>0$. This significantly simplifies the evaluation of the upper bound for large $L$ and $K$.
\end{remk}

Using (\ref{Eq:ML_Sol}), $X$ can be rewritten as 
\begin{IEEEeqnarray}{lll} \label{Eq:X_weighted}
X = \sum_{k=1}^{K} r[k] \mathsf{ln}\left( \frac{1+ \hat{s}[k] \mathsf{SIR}}{1+ s[k] \mathsf{SIR}}  \right)
\triangleq \sum_{k=1}^{K} \varpi[k]r[k],
\end{IEEEeqnarray}
which is basically a weighted sum of the observations. Note that given $\mathbf{s}$, $r[k],\,\,\forall k$, is a Poisson RV with mean $\lambda[k]=s[k]\bar{c}_{\mathrm{s}}+\bar{c}_{\mathrm{n}}$ and MFG $G_{r[k]}(t)=\mathsf{exp}(\lambda[k](e^{t}-1))$. Exploiting the properties of MGFs, namely $G_{aX}(t)=G_X(at)$, where $a$ is a constant, and $G_{X+Y}(t)=G_{X}(t)G_Y(t)$ where $X$ and $Y$ are independent RVs, we obtain
\begin{IEEEeqnarray}{lll} \label{Eq:MGF}
G_X(t)=\prod_{k=1}^{K} G_{r[k]}\left(\varpi[k]t\right) = \mathsf{exp}\left(\sum_{k=1}^K \lambda[k]\left(e^{\varpi[k]t}-1\right)\right). \quad\,\,\,
\end{IEEEeqnarray}
 The above result leads to the upper bound in (\ref{Eq:UpperGen}) and concludes the proof.

\section{Proof of Corollary~\ref{Corol:CER_UppBin}}\label{App:Corol_CER_UppBin}

Using the PEP, the CER is upper bounded based on the union bound as follows
\begin{IEEEeqnarray}{lll} \label{Eq:UB_Bin}
P_e^{\mathrm{code}}(\bar{\boldsymbol{\omega}}|\bar{\mathbf{c}}) & \leq \sum_{\forall\mathbf{s}} \sum_{\forall\hat{\mathbf{s}}\neq \mathbf{s}} P(\mathbf{s}\to\hat{\mathbf{s}}) \mathsf{Pr}(\mathbf{s}) \nonumber \\
 & = \frac{1}{M} \sum_{\forall\mathbf{s}} \sum_{\forall\hat{\mathbf{s}}\neq \mathbf{s}}  \mathsf{Pr}\{X>0\}+0.5\mathsf{Pr}\{X=0\}, \quad
\end{IEEEeqnarray}
where $X= \Lambda^{\mathrm{ML}}(\hat{\mathbf{s}}) - \Lambda^{\mathrm{ML}}(\mathbf{s})$. RV $X$ can be simplified as
\begin{IEEEeqnarray}{lll} 
X = \sum_{k=1}^K (\hat{s}[k]-s[k]) r[k] = \overset{X_2}{\overbrace{\sum_{k\in\widehat{\mathcal{K}}} \hat{s}[k]r[k]}}
 - \overset{X_1}{\overbrace{\sum_{k\in\mathcal{K}} s[k]r[k]}}, \quad
\end{IEEEeqnarray}
where $\mathcal{K}=\{k|s[k]=1\,\,\text{and}\,\,s[k]\neq\hat{s}[k]\}$ and $\widehat{\mathcal{K}}=\{k|\hat{s}[k]=1\,\,\text{and}\,\,s[k]\neq\hat{s}[k]\}$. Here, $X_1$ and $X_2$ are two \textit{independent} Poisson RVs with means $\lambda_1=\frac{d_{ij}(\bar{c}_{\mathrm{s}}+\bar{c}_{\mathrm{n}})}{2}$ and $\lambda_2=\frac{d_{ij}\bar{c}_{\mathrm{n}}}{2}$, respectively. Therefore, $X$ follows a Skellam distribution whose PDF is given in (\ref{Eq:Skellam}) \cite{Skellam}. Moreover,  since, for a given $\mathbf{s}$ and $\hat{\mathbf{s}}$, the Skellam distribution is a function of the Hamming distance $d_{ij}$, we can replace the summations in (\ref{Eq:UB_Bin}) by the summation over all $d_{ij}$ as in (\ref{Eq:CER_UppBin}). This completes the proof.

\section{Proof of Proposition~\ref{Prop:UppBinFull}}\label{App:PropUppBinFull} 
Let $\hat{\mathbf{s}}$ denote the detected codeword using the optimal detector. We divide the received vector $\mathbf{r}$ into two vectors $\tilde{\mathbf{r}}=[\tilde{r}_1,\tilde{r}_2,\dots,\tilde{r}_{\omega}]^T$ and $\hat{\mathbf{r}}=[\hat{r}_1,\hat{r}_2,\dots,\hat{r}_{K-\omega}]^T$ which correspond to the positions of the ``1"s and ``0"s in the transmitted codeword $\mathbf{s}$, respectively. Hereby, conditioned on $\mathbf{s}$, elements $\tilde{r}_i$ and $\hat{r}_j$ are independent Poisson RVs with means $\bar{c}_{\mathrm{s}}+\bar{c}_{\mathrm{n}}$ and $\bar{c}_{\mathrm{n}}$, respectively. Let us define $X=\mathsf{min}\{\tilde{r}_1,\tilde{r}_2,\dots,\tilde{r}_{\omega}\}$ and $Y=\mathsf{max}\{\hat{r}_1,\hat{r}_2,\dots,\hat{r}_{K-\omega}\}$. For the optimal detector and a full binary CW code, the CER is bounded as
\begin{IEEEeqnarray}{rll} \label{Eq:PoissMinMax}
\mathsf{Pr}\{X < Y \} \leq P_e^{\mathrm{code}}(K,\omega |\bar{\mathbf{c}})  \leq \mathsf{Pr}\{X \leq Y \}.
\end{IEEEeqnarray}
In fact, for events when $X=Y$ occurs, the detector selects with equal probability one of the hypotheses yielding the same value of $\Lambda^{\mathrm{ML}}(\mathbf{s})$. For the upper bound, we treat event $X=Y$ as an error and for the lower bound, we treat it as a correct decision. Using order statistics theory \cite{ISWCS2015IEEE,Order_Statistics}, the cumulative density function (CDF) of $X$ and the PDF of $Y$ are given by (\ref{Eq:PDFMaxMin}a) and (\ref{Eq:PDFMaxMin}b), respectively,  where $f_\mathcal{P}(\cdot,\lambda)$ and $F_{\mathcal{P}}(\cdot,\lambda)$ are in fact the PDF and CDF of a Poisson RV with mean $\lambda$, respectively \cite{Order_Statistics}. Using $F_X(x)$ and $f_Y(y)$, the lower and upper bounds in (\ref{Eq:PoissMinMax}) are given in (\ref{Eq:CER_sum}). This completes the proof.

\section*{Acknowledgment}
The authors would like to thank Prof. Andrea Goldsmith for her valuable suggestions and comments for an earlier version of this~paper.

\bibliographystyle{IEEEtran}
\bibliography{Ref_06_06_2017}

\begin{thebibliography}{10}
\providecommand{\url}[1]{#1}
\csname url@samestyle\endcsname
\providecommand{\newblock}{\relax}
\providecommand{\bibinfo}[2]{#2}
\providecommand{\BIBentrySTDinterwordspacing}{\spaceskip=0pt\relax}
\providecommand{\BIBentryALTinterwordstretchfactor}{4}
\providecommand{\BIBentryALTinterwordspacing}{\spaceskip=\fontdimen2\font plus
\BIBentryALTinterwordstretchfactor\fontdimen3\font minus
  \fontdimen4\font\relax}
\providecommand{\BIBforeignlanguage}[2]{{%
\expandafter\ifx\csname l@#1\endcsname\relax
\typeout{** WARNING: IEEEtran.bst: No hyphenation pattern has been}%
\typeout{** loaded for the language `#1'. Using the pattern for}%
\typeout{** the default language instead.}%
\else
\language=\csname l@#1\endcsname
\fi
#2}}
\providecommand{\BIBdecl}{\relax}
\BIBdecl

\bibitem{ISIT17_IEEE}
V.~Jamali, A.~Ahmadzadeh, N.~Farsad, and R.~Schober, ``{SCW Codes for Optimal
  CSI-Free Detection in Diffusive Molecular Communications},'' \emph{Accepted
  for presentation at IEEE ISIT}, Jun. 2017.

\bibitem{Nariman_Survey}
N.~Farsad, H.~Yilmaz, A.~Eckford, C.~Chae, and W.~Guo, ``{A Comprehensive
  Survey of Recent Advancements in Molecular Communication},'' \emph{IEEE
  Commun. Surveys Tutorials}, vol.~18, no.~3, pp. 1887--1919, third quarter
  2016.

\bibitem{Survey_Mol_Net}
T.~Nakano, M.~Moore, F.~Wei, A.~Vasilakos, and J.~Shuai, ``{Molecular
  Communication and Networking: Opportunities and Challenges},'' \emph{IEEE
  Trans. NanoBiosci.}, vol.~11, no.~2, pp. 135--148, Jun. 2012.

\bibitem{CellBio}
B.~Alberts, D.~Bray, K.~Hopkin, A.~Johnson, J.~Lewis, M.~Raff, K.~Roberts, and
  P.~Walter, \emph{{Essential Cell Biology}}.\hskip 1em plus 0.5em minus
  0.4em\relax New York, NY: Garland Science, 4th ed., 2014.

\bibitem{BioPhysic}
P.~Nelson, \emph{{Biological Physics: Energy, Information, Life}}.\hskip 1em
  plus 0.5em minus 0.4em\relax Freeman, 1st ed., 2008.

\bibitem{Survey_Mol_Nono}
I.~Akyildiz, F.~Brunetti, and C.~Blazquez, ``{Nanonetworks: A New Communication
  Paradigm},'' \emph{Comput. Net.}, vol.~52, pp. 2260--2279, Apr. 2008.

\bibitem{HamidJSAC}
R.~Mosayebi, H.~Arjmandi, A.~Gohari, M.~Nasiri-Kenari, and U.~Mitra,
  ``{Receivers for Diffusion-Based Molecular Communication: Exploiting Memory
  and Sampling Rate},'' \emph{IEEE J. Sel. Areas Commun.}, vol.~32, no.~12, pp.
  2368--2380, Dec. 2014.

\bibitem{TCOM_MC_CSI}
V.~Jamali, A.~Ahmadzadeh, C.~Jardin, C.~Sticht, and R.~Schober, ``{Channel
  Estimation for Diffusive Molecular Communications},'' \emph{IEEE Trans.
  Commun.}, vol.~64, no.~10, pp. 4238--4252, Oct. 2016.

\bibitem{NanoCOM16}
V.~Jamali, N.~Farsad, R.~Schober, and A.~Goldsmith, ``{Non-Coherent
  Multiple-Symbol Detection for Diffusive Molecular Communications},'' in
  \emph{Proc. ACM NanoCom}, Sept. 2016.

\bibitem{Berg}
H.~C. Berg, \emph{{Random Walks in Biology}}.\hskip 1em plus 0.5em minus
  0.4em\relax Princeton University Press, 1993.

\bibitem{ArmanMobileMC}
M.~Ahmadzadeh, V.~Jamali, A.~Noel, and R.~Schober, ``{Diffusive Mobile
  Molecular Communications Over Time-Variant Channels},'' \emph{IEEE Commun.
  Lett.}, 2017.

\bibitem{ArmanMCStat}
\BIBentryALTinterwordspacing
M.~Ahmadzadeh, V.~Jamali, and R.~Schober, ``{Statistical Analysis of
  Time-Variant Channels in Diffusive Mobile Molecular Communications},''
  \emph{Submitted to IEEE Globecom}, 2017. [Online]. Available:
  \url{https://arxiv.org/abs/1704.06298}
\BIBentrySTDinterwordspacing

\bibitem{Huber_Polar}
M.~Seidl, A.~Schenk, C.~Stierstorfer, and J.~B. Huber, ``{Polar-Coded
  Modulation},'' \emph{IEEE Trans. Commun.}, vol.~61, no.~10, pp. 4108--4119,
  Oct. 2013.

\bibitem{Ungerboeck_CodedMod}
G.~Ungerboeck, ``{Channel Coding with Multilevel/Phase Signals},'' \emph{IEEE
  Trans. Inf. Theory}, vol.~28, no.~1, pp. 55--67, Jan. 1982.

\bibitem{ConsCIR}
M.~Mahfuz, D.~Makrakis, and H.~Mouftah, ``{A Comprehensive Study of
  Sampling-Based Optimum Signal Detection in Concentration-Encoded Molecular
  Communication},'' \emph{IEEE Trans. NanoBiosci.}, vol.~13, no.~3, pp.
  208--222, Sept. 2014.

\bibitem{OOK_MC}
M.~U. Mahfuz, D.~Makrakis, and H.~T. Mouftah, ``{On the Characterization of
  Binary Concentration-Encoded Molecular Communication in Nanonetworks},''
  \emph{Nano Commun. Net.}, vol.~1, no.~4, pp. 289--300, Dec. 2010.

\bibitem{Nariman_Timing}
N.~Farsad, Y.~Murin, A.~Eckford, and A.~Goldsmith, ``{On the Capacity of
  Diffusion-Based Molecular Timing Channels},'' in \emph{IEEE ISIT}, Jul. 2016,
  pp. 1023--1027.

\bibitem{PPM_Pieroborn}
N.~Garralda, I.~Llatser, A.~Cabellos-Aparicio, E.~Alarc{\'o}n, and M.~Pierobon,
  ``{Diffusion-Based Physical Channel Identification in Molecular
  Nanonetworks},'' \emph{Nano Commun. Net., Elsevier}, vol.~2, no.~4, pp.
  196--204, Dec. 2011.

\bibitem{Huber_MultiLevCode}
U.~Wachsmann, R.~F.~H. Fischer, and J.~B. Huber, ``{Multilevel Codes:
  Theoretical Concepts and Practical Design Rules},'' \emph{IEEE Trans. Inf.
  Theory}, vol.~45, no.~5, pp. 1361--1391, Jul. 1999.

\bibitem{WeightCodeClass}
P.~R.~J. Ostergard, ``{Classification of Binary Constant Weight Codes},''
  \emph{IEEE Trans. Inf. Theory}, vol.~56, no.~8, pp. 3779--3785, Aug. 2010.

\bibitem{CW_qray}
Y.~M. Chee and S.~Ling, ``{Constructions for $q$-Ary Constant-Weight Codes},''
  \emph{IEEE Trans. Inf. Theory}, vol.~53, no.~1, pp. 135--146, Jan. 2007.

\bibitem{BalancedCode_Knuth}
D.~Knuth, ``{Efficient Balanced Codes},'' \emph{IEEE Trans. Inf. Theory},
  vol.~32, no.~1, pp. 51--53, Jan. 1986.

\bibitem{MutiplyCW}
Z.~Cherif, J.~L. Danger, S.~Guilley, J.~L. Kim, and P.~Solé, ``{Multiply
  Constant Weight Codes},'' in \emph{IEEE ISIT}, Jul. 2013, pp. 306--310.

\bibitem{MPPM_Optic}
H.~Sugiyama and K.~Nosu, ``{MPPM: A Method for Improving the Band-Utilization
  Efficiency in Optical PPM},'' \emph{J. Lightwave Technol.}, vol.~7, no.~3,
  pp. 465--472, March 1989.

\bibitem{MPPM_Rate}
C.~N. Georghiades, ``{Modulation and Coding for Throughput-Efficient Optical
  Systems},'' \emph{IEEE Trans. Inf. Theory}, vol.~40, no.~5, pp. 1313--1326,
  Sept. 1994.

\bibitem{Yilmaz_Poiss}
H.~B. Yilmaz and C.~B. Chae, ``{Arrival Modelling for Molecular Communication
  via Diffusion},'' \emph{Electron. Lett.}, vol.~50, no.~23, pp. 1667--1669,
  Nov. 2014.

\bibitem{Adam_Enzyme}
A.~Noel, K.~Cheung, and R.~Schober, ``{Improving Receiver Performance of
  Diffusive Molecular Communication with Enzymes},'' \emph{IEEE Trans.
  NanoBiosci.}, vol.~13, no.~1, pp. 31--43, Mar. 2014.

\bibitem{Acid_Base}
N.~Farsad and A.~Goldsmith, ``{A Molecular Communication System Using Acids,
  Bases and Hydrogen Ions},'' in \emph{IEEE SPAWC}, Jul. 2016, pp. 1--6.

\bibitem{CL_MF_IEEE}
\BIBentryALTinterwordspacing
V.~Jamali, M.~Ahmadzadeh, and R.~Schober, ``{On the Design of Matched Filters
  for Molecule Counting Receivers},'' \emph{IEEE Commun. Lett.}, 2017.
  [Online]. Available: \url{https://arxiv.org/abs/1705.01733}
\BIBentrySTDinterwordspacing

\bibitem{SortingComplexity}
P.~van Emde~Boas, ``{Preserving Order in a Forest in Less Than Logarithmic
  Time},'' in \emph{16th Ann. Symp. Found. Comput. Sci.}, 1975, pp. 75--84.

\bibitem{TableIntegSerie}
I.~S. Gradshteyn and I.~M. Ryzhik, \emph{Table of Integrals, Series, and
  Products}.\hskip 1em plus 0.5em minus 0.4em\relax 7th ed. Academic, 2007.

\bibitem{CodeMinDis}
C.~S. Laih and C.~N. Yang, ``{Design of Efficient Balanced Codes with Minimum
  Distance 4},'' \emph{IEE Proc. Commun.}, vol. 143, no.~4, pp. 177--181, Aug.
  1996.

\bibitem{StochGallager}
R.~G. Gallager, \emph{Stochastic Processes, Theory for Applications}.\hskip 1em
  plus 0.5em minus 0.4em\relax Cambridge, UK: Cambridge University Press, 2013.

\bibitem{Skellam}
J.~G. Skellam, ``{The Frequency Distribution of the Difference Between Two
  Poisson Variates Belonging to Different Populations.}'' \emph{J. Royal
  Statistical Society}, vol. 109, no. Pt 3, pp. 296--296, 1945.

\bibitem{ISWCS2015IEEE}
V.~Jamali, D.~S. Michalopoulos, M.~Uysal, and R.~Schober, ``{Outage Analysis of
  $q$-Duplex RF/FSO Relaying},'' in \emph{Proc. IEEE ISWCS (Invited Paper)},
  Brussels, Aug. 2015, pp. 1--5.

\bibitem{Order_Statistics}
H.~David and H.~Nagaraja, \emph{{Order Statistics}}, ser. Wiley Series in
  Probability and Statistics.\hskip 1em plus 0.5em minus 0.4em\relax Wiley,
  2004.

\end{thebibliography}

\end{document}